\documentstyle[12pt]{article}

\newcommand{\half}{\mbox{$\frac{1}{2}$}}
\newcommand{\eps}{\epsilon}
\renewcommand{\S}{s}
\newcommand{\gtlt}{\stackrel{>}{ <}}
\newcommand{\glN}{{\rm gl}_N}
\newcommand{\GLN}{{\rm GL}_N}
\newcommand{\cG}{{\cal G}}
\newcommand{\cH}{{\cal H}}
\newcommand{\id}{I}

\newcommand{\cP}{{\cal P}}

\newfam\msbfam
\batchmode\font\twelvemsb=msbm10 scaled\magstep1 \errorstopmode
\ifx\twelvemsb\nullfont\def\Bbb{\bf}
\else   \catcode`\@=11
        \font\tenmsb=msbm10 \font\sevenmsb=msbm7 \font\fivemsb=msbm5
        \textfont\msbfam=\tenmsb
        \scriptfont\msbfam=\sevenmsb \scriptscriptfont\msbfam=\fivemsb
        \def\Bbb{\relax\ifmmode\expandafter\Bbb@\else
                \expandafter\nonmatherr@\expandafter\Bbb\fi}
        \def\Bbb@#1{{\Bbb@@{#1}}}
       \def\Bbb@@#1{\fam\msbfam\relax#1}
        \catcode`\@=\active
\fi
\newcommand{\R}{{\Bbb R}}
\newcommand{\C}{{\Bbb C}}
\newcommand{\Z}{{\Bbb Z}}
\newcommand{\N}{{\Bbb N}}

\begin{document}
\begin{flushright}
June 25, 1999
\end{flushright}
\vspace{.4cm}

\begin{center}
{\Large \bf Finding and solving Calogero-Moser type
systems using Yang-Mills gauge theories}\\
\vspace{1 cm}
{\large Jonas Blom$^*$ and 
Edwin Langmann$^{*,**,***}$}
\vspace{0.3 cm}

$^*$
{\em Theoretical Physics, KTH, S-100 44 Stockholm, Sweden} \\
$^{**}$
{\em Dept.\ of Theoretical Physics, UU,  
Box 803, S-751 08 Uppsala, Sweden }\\
$^{***}${Dept.\ of Statistics, SLU, Box 7070, S-750 07 Uppsala, Sweden}
\end{center}

\begin{abstract}
Yang-Mills gauge theory models on a cylinder coupled to external
matter charges provide powerful means to find and solve certain
non-linear integrable systems. We show that, depending on the choice
of gauge group and matter charges, such a Yang-Mills model is
equivalent to trigonometric Calogero-Moser systems and certain known
spin generalizations thereof. Choosing a more general ansatz for the
matter charges allows us to obtain and solve novel integrable systems.
The key property we use to prove integrability and to solve these
systems is gauge invariance of the corresponding Yang-Mills model.

\end{abstract}

\section{Introduction}
\label{intro}

In a previous paper \cite{BL} we presented a novel class of integrable
spin--particle systems which contains known integrable systems of
Calogero-Moser (CM) type \cite{CM} and certain known spin generalization
\cite{W,Woj} thereof as special cases, and many other generalizations which
(to our knowledge) were not known before. Our method not only allowed
us to find and prove integrability of these models but also to solve
them explicitly. Subsequently two alternative derivations of these
models also proving integrability were given by Polychronakos
\cite{P1,P2}.

In the present paper we give a more detailed account of our approach
and also present extensions of our previous results. We made some
effort to make the paper easily accessible to different readers: those
who quickly want to get the flavor of how our method work, but also
those who are interested in the details. In the rest of this section
we give some introduction to CM type systems, describe our
method, and then give a plan for the paper.

About thirty years ago it was discovered that a dynamical system of
particles on the line described by the Hamiltonian
\begin{equation}
   \label{CMS}
 {\cal H}= \frac{1}{2}\sum_{\alpha = 1}^{N}  (p^{\alpha})^{2} +
 \frac{e^{2}}{2}\sum_{\stackrel{\alpha,\beta=1}{\alpha\neq \beta}}^{N}
 v(q^{\alpha}-q^{\beta})
\end{equation}
is completely integrable when the pair potential $v(r)$ equals a
Weierstrass elliptic function $\wp(r)$, important special cases of
which are $1/r^2$, $a^2/\sin^2(ar)$, and $a^2/\sinh^2(ar)$ \cite{CM}
(for an early review see \cite{OPcl}). Subsequently these models have
received much interest in different contexts, and various
generalizations of these models have been found and studied (for
recent reviews see, e.g., \cite{Prev,HPrev}). Recently it was
observed that these models can be obtained from Yang-Mills gauge
theories on the cylinder coupled to particular non-dynamical matter
charges \cite{gauge}.\footnote{This is actually implicit already in
earlier work; see, e.g.,\ \cite{LS1}} Exploring this relation further
we found and solved a large class of novel integrable spin--particle
systems \cite{BL}. In this paper we give a more detailed account of
these and some new results. The key property we use to prove
integrability and to solve these systems is gauge invariance of the
corresponding Yang-Mills model. This allows us to use different
gauges, i.e., to impose different constraints compatible with gauge
invariance. There is a gauge in which the dynamics of such a
Yang-Mills model is equivalent to the dynamics of a CM type system,
whereas in another gauge the dynamics is free and the solution can be
found trivially. Thus the solution of the former system can be
obtained from the latter solution by constructing a certain gauge
transformation.

Relating a CM type system to a Yang-Mills theory makes
integrability obvious and the construction of conservation laws
(nearly) trivial. Moreover, this relation provides a rather simple
method to construct an explicit solution. (This method can be regarded
as an extension of the projection method \cite{OPcl}.) Since
there is a large freedom in choosing the external matter charges, one
can obtain and solve a large number of different
integrable systems. We believe that it should be possible to find
other integrable models using our method, for example by considering
more general gauge groups etc.

The plan of this paper is as follows.  In Section \ref{sec_YM} we 
summarize the facts about Yang-Mills gauge theories on a cylinder 
which we need in the sequel.  The derivation of a certain class of 
dynamical systems from these gauge theories is explained in Section 
\ref{dynsys}.  We present two different arguments: the first argument 
(Section \ref{canon}) is quick but only heuristic, and the second 
(Section \ref{diffdyn}) is somewhat less intuitive but rigorous.  In 
Section \ref{sec_intsys} we show how to exploit the relation of these 
dynamical systems to gauge theories to explicitly solve interesting 
special cases of these systems, and we also show how to obtain Lax 
pairs and conservation laws in our formalism.  As a warm--up, we first 
show how known results about CM models with potentials $1/r^2$, 
$a^2/\sin^2(ar)$ and $a^2/\sinh^2(ar)$ can be recovered (Sections 
\ref{sec_simpex}--\ref{suthsh}).  We will also refer to the former 
case as {\em Calogero model} and the latter as {\em Sutherland model}.  
We then derive and extend the results for the novel systems found in 
\cite{BL} (Section \ref{general}).  A novel class of models which can 
not be solved in such an explicit manner but still should be 
integrable is discussed in Section \ref{general1}.

Readers who only want to get a flavor of how our methods work are
advised to read the beginning of Section \ref{sec_intsys} and proceed
to Sections \ref{sutherland} and \ref{general}.

\vspace*{0.5cm} \noindent {\bf Notation.} We denote as $\glN$ the
complex $N\times N$ matrices, $\GLN$ the complex invertible $N\times
N$ matrices, and $\id$ the $N\times N$ unit matrix.

\section{Yang-Mills theory on a cylinder}
\label{sec_YM}
In this section we fix notation and define the gauge theory models of
interest for our purposes. We also discuss different gauge conditions
which will play important roles in the sequel.

\subsection{Definition}
We consider Yang-Mills theory on a cylinder with external matter 
sources.  We restrict ourselves to a two-dimensional spacetime which 
is a cylinder, i.e., the time coordinate is $t\in \R$, and space is a 
circle parameterized by $x\in [-\pi,\pi]$.  In the following, 
$\mu,\nu\in \{0,1\}$ are spacetime indices, and repeated spacetime 
indices are summed over.\footnote{Note that we will use this summation 
convention {\em only} for spacetime indices.}
Our metric tensor is ${\rm diag}(1,-1)$.

Our starting point is the Lagrangian
\begin{equation}
{\cal L}(t,x)=-\frac{1}{2\pi}\left(\frac{1}{4} tr\left[F_{\mu
         \nu}(t,x)F^{\mu\nu}(t,x) + A_{\nu}(t,x)j^{\nu}(t,x)\right]
         \right)
         \label{lagrangian}
\end{equation}
where we set 
\begin{equation}
j^1\equiv 0,\quad j^0\equiv \rho .  \label{la2}
\end{equation}
This Lagrangian describes a dynamical Yang-Mills field $A_{\nu}$ coupled
to an external matter current $j^{\nu}$ whose spatial component
vanishes. We shall denote the temporal component $j^0=\rho$ of this
matter current as {\em charge}. The Yang-Mills curvature $F_{\mu\nu}$
is defined as
\begin{equation}
  F_{\mu\nu}=\partial_{\mu}A_{\nu}-\partial_{\nu}A_{\mu}
  + ig[A_{\mu},A_{\nu}].
\end{equation}
Note that the only non-trivial component of this is
\begin{equation}
\label{E}
E(t,x):=\, F_{01}(t,x)=-F_{10}(t,x).
\end{equation}
We assume a representation of the structure groups such that
$A_\mu$ and $\rho$ are $\glN$-valued functions.

\subsection{Gauges}
\label{gauges}
The Lagrangian defined in Eqs.\ (\ref{lagrangian}) and (\ref{la2}) 
is obviously invariant under all transformations
\begin{eqnarray}
   \label{gauge_trafo}
A_\mu(t,x)&\to& A_\mu^U(t,x) = U^{-1}(t,x) A_\mu(t,x) U(t,x) + 
\frac{1}{ig} U^{-1}(t,x)\partial_\mu U(t,x)
\nonumber \\ \nopagebreak \rho(t,x)&\to & 
\rho^U(t,x)= U^{-1}(t,x) \rho(t,x) U(t,x)
\nonumber \\ \nopagebreak E(t,x) &\to& E^U(t,x)= 
U^{-1}(t,x) E(t,x) U(t,x) 
\end{eqnarray} 
for all differentiable $\GLN$-valued 
functions $U$ on spacetime (the last relation is of course not 
independent since it follows from the first).  This implies that the 
equations of motion derived from the Lagrangian 
(\ref{lagrangian})--(\ref{la2}) are also invariant under all 
these transformations. We will consider 
functions $U(t,x)$ which are differentiable only {\em locally} in 
time, i.e., it is differentiable for $t$ in some non-empty interval 
$(t_1,t_2)$ and all $x\in [-\pi,\pi]$.  We denote the set of all such 
functions as $\cG_{(t_1,t_2)}$.  We refer to the transformations in 
Eq.\ (\ref{gauge_trafo}) as {\em gauge transformations}.  {\em Gauge 
invariance}, i.e.,\ invariance under all gauge transformations, will 
allow us to impose additional constraints on the gauge fields.  We 
call these constraints {\em gauges}.  There are two different gauges 
which will be important for us.

\subsubsection*{Diagonal Coulomb Gauge}
This gauge is defined by the condition
that $A_1$ is a diagonal matrix which is independent of $x$, i.e.,
\begin{equation}
\label{DCG}
A_1(t,x)= Q(t) = {\rm diag}\left( q^1(t), q^2(t), \ldots, q^N(t)
\right).
\end{equation}
To show that this indeed is a gauge one has to prove that it is always
possible to find a gauge transformation bringing the Yang-Mills field
into the form of Eq.\ (\ref{DCG}), i.e.,\ for each (generic)
Yang-Mills configuration $A_1(t,x)$ one can find a gauge
transformation $U$ such that $A_1^U(t,x)$ defined by Eq.\
(\ref{gauge_trafo}) has the form of $Q(t)$ in Eq.\ (\ref{DCG}). For
completeness we now recall a proof of this by explicit construction
of the required $U$ (see \cite{BL} and references therein).

We first note that the condition $A_1^U=Q$ is equivalent to the
differential equation $\partial_1 U +ig A_1 U= ig U Q $. To solve this
equation we first consider the boundary value problem 
\begin{equation}
\label{S_def}
\partial_1 S(t,x) + ig A_1(t,x) S(t,x) =0, \quad  S(t,-\pi)=\id . 
\end{equation}
This problem has a unique solution (see, e.g., Ref.\ \cite{pri}, Theorems
1.1 and 2.1) which we write as
\begin{equation}
\label{S}
  S(t,x)= \cP \exp\left( -ig \int_{-\pi}^x dy \, A_1(t,y) \right)
\end{equation}
where the symbol $\cP\exp$ denotes the path ordered
exponential.\footnote{in the terminology of Ref.\ \cite{pri}, $S(t,x)$
is identical with the product integral $\prod_{-\pi}^x e^{-ig A_1(t,s)
ds}$ } It is important to note that $S(t,x)$ is not a gauge
transformation since it is not periodic in $x$ (its values at $x=-\pi$
and $\pi$ are different in general). However, it can be used to
construct a gauge transformation as follows. We first introduce an
important technical condition. We call a gauge field $A_1(t,x)$ {\em
regular} (at time $t$) if the corresponding matrix $S(t,\pi)$ is
non-degenerate, i.e., all the eigenvalues of $S(t,\pi)$ are different.
If $A_1(t,x)$ is regular then there exists an invertible matrix $V(t)$
diagonalizing $S(t,\pi)$ (see e.g.\ Theorem 10.2.4. in \cite{Mir}),
i.e.,
\begin{equation}
  \label{VSV}
  V(t)^{-1} S(t,\pi) V(t) = e^{-ig  2\pi Q(t) }
\end{equation}
for some diagonal matrix $Q(t)={\rm diag}(q^1(t),\ldots,q^N(t))$.
We now claim that the function
\begin{equation}
  \label{U}
  U(t,x) = S(t,x) V(t) e^{ig (x+\pi) Q(t)}
\end{equation}
is periodic. Indeed,
$$
U(t,\pi) = V(t) V(t)^{-1} S(t,\pi) V(t) e^{ig 2 \pi Q(t)} = V(t)
=U(t,-\pi)
$$
where we inserted $V(t)V(t)^{-1}=\id$ and used Eq.\ (\ref{VSV}).
Moreover, $U$ Eq.\ (\ref{U})
satisfies
$\partial_1 U +ig A_1 U= ig U Q $
equivalent to $A_1^U=Q$.
It is also easy to see that $U(t,x)$ is invertible (see e.g.\ Ref.\
\cite{pri}, Section 1.1). We are left to show that $U(t,x)$
is a differentiable function on spacetime. This might seem
trivial but is actually not since $V(t)$ and $Q(t)$
as defined in Eq.\ (\ref{VSV})
can be discontinuous in $t$; see, e.g., \cite{Blau}.
However, if $A_1(t,x)$ is regular for all $t$
in an open time interval $(t_1,t_2)$ then $Q(t)$ and $V(t)$
can be chosen to be differentiable in this time interval
\cite{Blau}, and $U(t,x)$ in Eq.\ (\ref{U})
is indeed a differentiable function, i.e., a
gauge transformation in $\cG_{(t_1,t_2)}$.
We call a Yang-Mills field $A_1(t,x)$ {\em generic}
if it is regular for all times $t\in\R$.\\

It is worth noting that the function $V(t)$ is not
unique: all transformations
\begin{equation}
\label{residual}
V(t)\to V'(t)= V(t) D(t), \quad D(t)= {\rm diag}(d^1(t),\ldots, d^N(t))
\end{equation} 
with arbitrary differentiable and non-zero functions
$d^\alpha(t)$, $\alpha=1,2,\ldots, N$, are compatible with the
conditions determining $V(t)$. This corresponds to the residual gauge
freedom which remains after imposing the diagonal Coulomb gauge: gauge
transformations Eq.\ (\ref{gauge_trafo}) with $U(t,x)=D(t)$ (diagonal
and independent of $x$) leave the diagonal Coulomb gauge Eq.\
(\ref{DCG}) invariant but act non-trivially on $A_0$ and $\rho$.

We note that the Eqs.\ (\ref{S_def}) and (\ref{VSV}) provide the recipe how
to compute $Q(t)$ from a given Yang-Mills configuration $A_1(t,x)$.
This will play an important role for us in the sequel.

\vspace*{0.3cm}\noindent {\em Remark:} 
We note that in our examples later the
eigenvalues $e^{-ig 2\pi q^\alpha(t)}$ of $S(t,\pi)$ have the
following physical interpretation: $q^\alpha(t),\; \alpha=1,\ldots,N$,
correspond to the positions of interacting particles in a dynamical
system. The technical condition of a Yang-Mills field being regular
in some interval thus means that the particles do not collide with
each other. Since the particle interactions are repulsive in all
dynamical systems we encounter, it is plausible that regularity holds
for all times, but we will be able to prove this only in certain
special cases. In general we can prove our results only locally in
time, i.e., for time intervals where no particle collisions occur.

\subsubsection*{Weyl gauge}
This gauge is defined by the condition
\begin{equation}
A_0(t,x)=0 .
\label{weyl}
\end{equation} 
To show that this is a gauge one can use a similar argument as
above, i.e.,\ for a given $A_0(t,x)$, the function 
\begin{equation} 
W(t,x)= \cP
\exp\left( -ig \int_{0}^t dt' \, A_0(t',x) \right), 
\end{equation} 
is such
that $A_0^W(t,x)$ Eq.\ (\ref{gauge_trafo}) vanishes. Note that $W(t,x)$
here is indeed a gauge transformation (i.e.,\ a periodic, invertible,
and differentiable function on spacetime). We note that if our
spacetime was a torus and not a cylinder, the Weyl- and the diagonal
Coulomb gauge would be more similar: one of them is obtained from the
other by interchanging the space- and time coordinates $x$ and $t$.
The reason why the Weyl gauge is simpler on the cylinder is that there
is no periodicity condition in $t$.

\section{Dynamical systems from Yang-Mills theories on the cylinder}
\label{dynsys}
In this section we show how to derive certain dynamical systems from
Yang-Mills systems on a cylinder with non-dynamical external matter
charges. We will use two different arguments which both lead to the
same results. In Section \ref{canon} we use a canonical procedure to
formulate our Yang-Mills system as an infinite dimensional Hamiltonian
system with a constraint called Gauss' law. By solving Gauss'
law in the diagonal Coulomb gauge Eq.\ (\ref{DCG}) we obtain an
integrable non-linear Hamiltonian system which generalizes the
Calogero- and Sutherland models discussed in Section \ref{intro}.
This Hamiltonian method is conceptually simple but not quite
satisfactory from a mathematical point of view. To make it
mathematically rigorous would require a deeper analysis which is
beyond the scope of the present paper. Instead, we discuss an
alternative and complimentary method in Section \ref{diffdyn}. This
method uses only the equations of motion and thus circumvents all the
mathematical difficulties which one would have to face in a rigorous
discussion of the method in Section \ref{canon}.

\subsection{Hamiltonian approach}
\label{canon}
In the following it is useful to denote the elements of the matrices
$M=A_\mu,E,\rho$ by $M^{\alpha\beta}$, $\alpha,\beta=1,2,\ldots N$,
and $tr$ the usual $N\times N$ matrix trace. Noting that
$tr(MM')=\sum_{\alpha,\beta=1}^N M^{\alpha\beta}(M')^{\beta\alpha}$
and following the standard canonical procedure \cite{Sun} we obtain
from the Lagrangian (\ref{lagrangian})--(\ref{la2}) 
(up to a surface term which
vanishes due to the periodicity of $A_0 E$ in $x$)
\begin{equation}
  {\cal H}=\frac{1}{2\pi}\int_{-\pi}^{\pi}dx
  \left[tr\left(\frac{1}{2}E^{2} -
      A_{0} [\partial_{1}E +ig[A_{1},E]-\rho  ] \right)\right] .
  \label{hamiltonian}
\end{equation}
The variable conjugate to $A_1^{\alpha\beta}(x)$ is
$E^{\beta\alpha}(x)$, which implies the Poisson brackets
\begin{equation}
  \label{canon_rel}
  \{A_{1}^{\alpha\beta}(x),E^{\alpha'\beta'}(y)\} = 2\pi
  \delta^{\alpha\beta'} \delta^{\beta\alpha'} \delta(x-y).
\end{equation}
Since the Lagrangian is independent of $A_0^{\alpha\beta}(x)$ its
conjugate momentum, which we denote as $\Pi_0^{\beta\alpha}(x)$, has
to vanish: $\Pi_0\simeq 0$. Here `$\simeq$' means that this is to be
regarded as a constraint \cite{Sun}. This primary constraint implies
the secondary constraint $\partial_0 \Pi_0 = \{ \Pi_0 ,H \}\simeq 0$,
i.e.,
\begin{equation}
G(x): =\, \partial_{1}E(x)+ig[A_{1}(x),E(x)]-\rho(x) \simeq 0,
\label{Gauss}
\end{equation}
which is called {\em Gauss' law}. The tertiary constraint $\partial_0 G(x)
= \{ G(x),H \} \simeq 0 $ together with $\{A_1(x),\rho(y)\}= \{
E(x),\rho(y)\} =0$ then fixes the Poisson bracket of the charges (see
Appendix \ref{sec_pb} for details)
\begin{equation}
  \label{rho_poi}
  \{\rho^{\alpha\beta}(x),\rho^{\alpha'\beta'}(y) \}
  \simeq ig2\pi\left[ \rho^{\alpha\beta'}(x) \delta^{\beta\alpha'} -
    \rho^{\alpha'\beta}(x) \delta^{\beta'\alpha} \right] \delta(x-y)
     .
\end{equation}
Note that,
$$
  \{G^{\alpha\beta}(x),G^{\alpha'\beta'}(y) \}
  \simeq ig2\pi\left[ G^{\alpha\beta'}(x) \delta^{\beta\alpha'} -
    G^{\alpha'\beta}(x) \delta^{\beta'\alpha} \right] \delta(x-y) , 
$$
and the Poisson brackets above also fix the time evolution of the
charges, $\partial_0{\rho}(x) = \{ \rho(x),H \}$.

We now exploit the gauge freedom and impose the gauges discussed in
the last section. We shall do that by replacing `$\simeq$' by `$=$'
in the equations given above, i.e., we ignore possible subtleties and
treat constraints like strict equalities. We therefore regard the
following as a simple but only heuristic argument, as discussed above.

\subsubsection*{Diagonal Coulomb Gauge}

As shown in the
last section, it is always possible to transform $A_1$ to a
$x$-independent diagonal matrix. This means that the Yang-Mills field
only has a finite number of true dynamical degrees of freedom. In
order to obtain a Hamiltonian for these degrees of freedom only, we
impose the gauge condition Eq.\ (\ref{DCG}). We then use Gauss' law
to eliminate all but the true dynamical degrees of freedom.

To do this we will use Fourier transformations
\begin{equation}
\hat E^{\alpha\beta} (n) = \int_{-\pi}^{\pi} dx\, e^{-i nx}
E^{\alpha\beta} (x), \quad n\in\Z 
\end{equation}
and similarly for $A_1$ and $\rho$. Using this and imposing Eq.\
(\ref{DCG}), the Gauss' law can
be rewritten in component form as
\begin{equation}
  \label{E_hat}
  i \left(n + g [q^{\alpha}-q^{\beta} ]\right) \hat
  E^{\alpha\beta}(n) =\hat\rho^{\alpha\beta}(n).
\end{equation}
For $\alpha=\beta$ and $n = 0$, (\ref{E_hat}) implies
\begin{equation}
  \label{rho_cons}
  \hat{\rho}^{\alpha\alpha}(0) = 0.
\end{equation}
In all other cases, Eq.\ (\ref{E_hat}) can be solved for
$\hat{E}^{\alpha\beta}(n)$. Inserting these in the Hamiltonian
(\ref{hamiltonian}), we obtain 
\begin{eqnarray*} 
{\cal
H}&=&\frac{1}{4\pi}\int_{-\pi}^{\pi} dx \; tr (E(x)^{2} ) =
\frac{1}{8\pi^2} \sum_{n \in \Z} \sum_{\alpha,\beta = 1}^N
\hat{E}^{\alpha\beta}(n)\hat{E}^{\beta\alpha}(-n) = 
\nonumber\\
&=&\sum_{\alpha = 1}^N \frac{\hat{E}^{\alpha\alpha}(0)
\hat{E}^{\alpha\alpha}(0)}{2(2\pi)^2} + \frac{1}{8\pi^2}\sum_{n \in
\Z} \sum_{\alpha, \beta=1}^N
(1-\delta_{n,0}\delta^{\alpha\beta})
\frac{\hat{\rho}^{\alpha\beta}(n)\hat{\rho}^{\beta\alpha}(-n)}
{(n+g[q^\alpha-q^\beta])^2}. 
\end{eqnarray*} 
Here a comment on our
notation is in order: for $\alpha=\beta$ and $n=0$ the last term
seems ambiguous since it formally is $0/0$. However, from
our derivation above it is clear that this has to be interpreted as
$0$, i.e., here and in the following
$$
(1-\delta_{n,0}\delta^{\alpha\beta})(\cdots)
\equiv 0 \quad \mbox{ if $n=0$ and $\alpha=\beta$ }
$$
even if $(\cdots)$ happens to be infinite.

We now find it convenient to introduce the following suggestive
notation,
\begin{equation}
  p^\alpha = \frac{\hat{E}^{\alpha\alpha}(0)}{2\pi}
  \label{p_alpha} .
\end{equation}
Moreover, we will see later that the functions
$\hat{A}_0^{\alpha\alpha}(t,0)\equiv a^\alpha(t)$ play a special role
and can be chosen arbitrarily. Thus despite of the constraint Eq.\
(\ref{rho_cons}) we can leave the corresponding term $\propto
\sum_\alpha a^\alpha(t) \hat{\rho}^{\alpha\alpha}(t,n=0)$ in the
Hamiltonian. We thus obtain 
\begin{equation} 
{\cal H} = \sum_{\alpha = 1}^N
\frac{(p^\alpha)^2}{2} + \frac{1}{8\pi^2}\sum_{n \in \Z}
\sum_{\alpha,\beta = 1}^N (1-\delta_{n,0}\delta^{\alpha\beta})
\frac{\hat{\rho}^{\alpha\beta}(n)\hat{\rho}^{\beta\alpha}(-n)}
{(n+g[q^\alpha-q^\beta])^2} + \frac{1}{4\pi^2} \sum_{\alpha=1}^N
a^\alpha \hat{\rho}^{\alpha\alpha}(0)
                \label{pq_hamiltonian}
\end{equation}
where $a^\alpha (t)$ play the role of an arbitrary external time
dependent field which contribute to the time evolution of
$\hat{\rho}^{\alpha\beta}(n)$.

Using Eq.\ (\ref{canon_rel}) it
is easy to verify that
\begin{equation}
\label{canonical}
  \{ q^\alpha,p^\beta \} = \left\{
   A_1^{\alpha\alpha}(x),\frac{\hat{E}^{\beta\beta}(0)}{2\pi}
  \right\} = \left\{A_1^{\alpha\alpha}(x),
    \frac{ \int_{-\pi}^{\pi}  E^{\beta\beta}(y) dy
      }{2\pi} \right\} = \delta^{\alpha\beta}.
\end{equation}
Hence, it is natural to interpret $p^\alpha$ as
particle momenta and $q^\alpha$ as the corresponding
position variables.
{}From the Poisson bracket (\ref{rho_poi}) we obtain
the corresponding Poisson bracket for $\hat{\rho}$,
\begin{equation}
  \label{rhohat_poi}
  \{\hat{\rho}^{\alpha \beta}(n),\hat{\rho}^{\alpha' \beta'}(m) \}
  = ig2\pi[\hat{\rho}^{\alpha\beta'}(n+m)\delta^{\beta\alpha'} -
  \hat{\rho}^{\alpha' \beta}(n+m)\delta^{\beta'\alpha}].
\end{equation}
We interpret the $\hat{\rho}^{\alpha\beta}(n)$ as spin degrees of freedom.
Hence, we have obtained
a dynamical system described by the
Hamiltonian (\ref{pq_hamiltonian}), together with the Poisson brackets
(\ref{canonical}) and (\ref{rhohat_poi}). All other Poisson brackets vanish.

It is interesting to note that the Hamiltonian Eq.\ (\ref{pq_hamiltonian})
is invariant under the transformations
\begin{equation}
  q^\alpha\to q^\alpha + \frac{m^\alpha}{g},\quad
  p^\alpha\to p^\alpha,\quad
  \hat{\rho}^{\alpha\beta}(n) \to
  \hat{\rho}^{\alpha\beta}(n+m^\alpha-m^\beta)
\end{equation}
for all integers $m^\alpha$. Thus, if $q^\alpha$,
$\alpha=1,\ldots,N$, are real, this model describes particles moving
on a circle of length $1/g$ and interacting with a potential whose
strength depends on the spin degrees of freedom. Moreover, this
Hamiltonian is invariant under all permutations of the particle labels.
It is worth noting that the existence of these symmetry
transformations is due to the Gribov ambiguities of the diagonal
Coulomb gauge as discussed in Ref.\ \cite{LS2}.

Using the Hamiltonian Eq.\ (\ref{pq_hamiltonian}) and the Poisson
brackets given above it is straightforward to derive the equations of
motion $\partial_0 X=\{X,\cH \}$ for $X=q^\alpha,p^\alpha$ and
$\hat\rho^{\alpha\beta}(n)$. (We will write down these equations in
Section \ref{diffdyn}.) Note that these equations are highly non-linear.

\subsubsection*{Weyl gauge}
In the Weyl gauge, the Hamiltonian is simply
\begin{equation}  
{\cal
H}=\frac{1}{4\pi} \int_{-\pi}^{\pi} dx\, tr (E(x)^{2} )
\label{weyl_hamiltonian}
\end{equation} 
which shows that we have a free (non-interacting) system.
Still, the system is not trivial since we have to account for Gauss'
law Eq.\ (\ref{Gauss}). However, it is easy to see that all quantities
$tr[E(x)^n]$ (arbitrary $x\in[-\pi,\pi]$ and $n\in\N$) are conserved
in time $t$. Thus there is an infinite number of conservation laws.
(Note that we do not claim that these conservation laws are
independent.) Since the system is free, we can actually solve the
system in the Weyl gauge (this will be done in the next section).

\subsection{Alternative approach based on equations of motion}
\label{diffdyn} Our discussion in the last section suggests that our
gauge theory is an integrable system and equivalent to a non-linear
dynamical system which is obtained by imposing the diagonal Coulomb
gauge. Instead of trying to make the argument in the last section
rigorous (which would be interesting but is beyond the scope of the
present paper) we now present an alternative, less intuitive, argument
leading to the same conclusions but avoiding the mathematical
difficulties of the canonical procedure.

We start with the Lagrangian equation obtained from the Lagrangian
(\ref{lagrangian}). They can be written as (recall that our metric
tensor used to rise and lower spacetime indices is ${\rm
diag}(1,-1)$, and that we use a summation convention for spacetime
indices $\mu,\nu\in \{ 0,1\}$)
\begin{equation}  
\label{123}
[D_\mu, F^{\mu\nu}]=j^\nu, \quad F_{\mu\nu}=\frac{1}{ig}[D_\mu,D_\nu]
\label{lageq}
\end{equation} 
with $j^1=0$ and $j^0=\rho$. Here we introduced the covariant 
derivative 
\begin{equation}  
 D_\mu=\partial_\mu + ig A_\mu . 
\end{equation} 
In this
paragraph (only!) we use a notation which takes into account the
Leibniz rule for differentiation. In this notation one
distinguishes, e.g., $(\partial_\nu A_\mu)$ from $\partial_\nu
A_\mu=(\partial_\nu A_\mu) + A_\mu \partial_\nu$ which is obtained by
using the Leibniz rule. Thus $(\partial_\nu
A_\mu)=[\partial_\nu,A_\mu]$. In this notation, gauge transformations
Eq.\ (\ref{gauge_trafo}) can be simply written as $D_\mu\to U^{-1}D_\mu
U$ and $j_\mu\to U^{-1}j_\mu U$, and it is thus obvious that the
Lagrangian equations above are gauge invariant. These equations imply
$$ 
[D_\nu,j^\nu]=[D_\nu,[D_\mu, F^{\mu\nu}]] =
ig[F_{\mu\nu},F^{\mu\nu}] + [D_\mu,[D_\nu, F^{\mu\nu}]] =
-[D_\mu,j^\mu] 
$$
where we used the Jacobi identity, the
anti-symmetry for the commutator, and $[F_{\mu\nu},F^{\mu\nu}]=0$
(cf.\ Eq.\ (\ref{E})). We thus also have the equation $[D_\nu , j^\nu]
=[D_0,\rho]=0$.

We can now forget that these differential equations originate from a
Lagrangian and instead use them as definition of our Yang-Mills model.
We find it convenient to write these equations in the following form
(we use the definition in Eq.\ (\ref{E}); to simplify notation we write
$(\partial_\mu X)$ as $\partial_\mu X$ in the sequel),
\begin{eqnarray}
\partial_0 A_1 = E + \partial_1 A_0 + ig  [A_1,A_0] \label{1} \\
\partial_0 E + ig  [A_0,E] = 0 \label{2} \\
\partial_0 \rho + ig  [A_0,\rho] = 0 \label{3}\\
\partial_1 E + ig  [A_1,E] = \rho  . \label{4}
\end{eqnarray}
Note that gauge invariance trivially implies the important

\vspace*{0.3cm}\noindent
{\bf Proposition:}
{\em If $A_\mu(t,x), E(t,x)$ and
$\rho(t,x)$ is a solution to the Eqs.\ (\ref{1})--(\ref{4}), then
$A_\mu^U(t,x), E^U(t,x)$ and
$\rho^U(t,x)$ Eq.\ (\ref{gauge_trafo}) is also a solution for arbitrary
differentiable $\GLN$-valued functions $U(t,x)$ on spacetime}. \\
\vspace*{0.3cm}

We now rewrite these equations in our two different gauges.

\subsubsection*{Diagonal Coulomb Gauge}
As argued in
Section \ref{sec_YM}, it is always possible to find a gauge
transformation (given by Eq.\ (\ref{U})) bringing $A_1$ into the form
Eq.\ (\ref{DCG}). We now impose this gauge, i.e.,\ we replace
$A_\mu,E$ and $\rho$ by $A_1^U, E^U$ and $\rho^U$ with $U$ given by
Eq.\ (\ref{U}). Using Fourier transformation, Eqs.
(\ref{1})--(\ref{4}) can be written as (to simplify notation, we still
use the symbols $E,A_0,\rho$ instead of $E^U,A_0^U,\rho^U$) 
\begin{equation}  
  \label{f1}
  2\pi\delta_{n,0}\delta^{\alpha\beta}\partial_0 q^\alpha(t) =
  \hat{E}^{\alpha\beta}(t,n) +
  i(n+g[q^\alpha(t)-q^\beta(t)])\hat{A}_0^{\alpha\beta}(t,n)  ,
\end{equation}
\begin{equation}  
\label{f2}
  \partial_0 \hat{E}^{\alpha\beta}(t,n) = -\frac{ig}{2\pi}\sum_{k\in\Z}
  \sum_{\gamma=1}^N \left(
  \hat A_0^{\alpha\gamma}(t,k) \hat E^{\gamma\beta}(t,n-k) -
  \hat E^{\alpha\gamma}(t,k) \hat A_0^{\gamma\beta}(t,n-k) \right) ,
\end{equation}
\begin{equation}  
  \label{f3}
  \partial_0 \hat{\rho}^{\alpha\beta}(t,n) = -\frac{ig}{2\pi}
  \sum_{k \in \Z}\sum_{\gamma=1}^N \left(
    \hat{A}_0^{\alpha\gamma}(t,k)\hat{\rho}^{\gamma\beta}(t,n-k) -
    \hat{\rho}^{\alpha\gamma}(t,k)\hat{A}_0^{\gamma\beta}(t,n-k) 
  \right) , 
\end{equation}
\begin{equation}  
  \label{f4}
  \hat{\rho}^{\alpha\beta}(t,n) =
  i(n+g[q^\alpha(t)-q^\beta(t)])\hat{E}^{\alpha\beta}(t,n) .
\end{equation}
Note that these equations determine all components of
$\hat{A}_0^{\alpha\beta}(t,n)$ except for $\alpha=\beta$ and $n=0$.
We will use the following notation for the unspecified components,
\begin{equation}  
\label{Aa}
\hat{A}_0^{\alpha\alpha}(t,0):=a^\alpha(t)
\label{aterm}
\end{equation}
which are arbitrary functions. They
correspond to the residual gauge freedom which is not fixed by
the diagonal Coulomb gauge, as discussed after Eq.\ (\ref{residual}).

Eq. (\ref{f1}) yields
\begin{equation}
  \label{E_exp}
  \hat{A}_0^{\alpha\beta} (t,n) =  a^\alpha(t)\delta^{\alpha\beta}\delta_{n,0}
  + \left(1-\delta^{\alpha\beta}\delta_{n,0}\right)
  \frac{i\hat{E}^{\alpha\beta} (t,n)}{\left(n + g
      [q^{\alpha}(t)-q^{\beta}(t) ]\right)},
\end{equation}
and Eq. (\ref{f4}) implies
\begin{equation}
\label{rhoexp}
\hat{E}^{\alpha\beta} (t,n) =
2\pi p^\alpha(t) \delta^{\alpha\beta}\delta_{n,0}
+\left(1-\delta^{\alpha\beta}\delta_{n,0} \right)
\frac{-i\hat\rho^{\alpha\beta} (t,n)}{
    \left(n + g [q^{\alpha}(t)-q^{\beta}(t) ]\right)}
\end{equation}
(we used the notation $\hat E^{\alpha\alpha} (t,0) = 2\pi
p^{\alpha}(t)$ introduced in Section \ref{canon}). By combining these
equations we obtain
\begin{equation}
  \label{A0_exp}
  \hat{A}_0^{\alpha\beta}(t,n) = a^\alpha(t)\delta^{\alpha\beta}\delta_{n,0}
  + \left(1-\delta^{\alpha\beta}\delta_{n,0}\right)
  \frac{\hat{\rho}^{\alpha\beta}(t,n)}{(n+g[q^{\alpha}(t)-q^{\beta}(t)])^2}.
\end{equation}

We now use these relations to derive the time evolution of our
dynamical quantities in the diagonal Coulomb gauge.
Putting $\beta = \alpha$ and $n=0$ in Eq.
(\ref{f1})
and using again $\hat E^{\alpha\alpha} (t,0)
= 2\pi p^{\alpha}(t)$
we obtain
\begin{equation}
\label{qdot}
\partial_0 q^{\alpha}(t)
= p^{\alpha}(t) .
\end{equation}
The time evolution of the $p^\alpha(t)=\hat E^{\alpha\alpha} (t,0)/2\pi$
follows by inserting Eqs.
(\ref{rhoexp}) and (\ref{A0_exp}) in Eq. (\ref{f2}):
\begin{equation}
\label{Pdot}
\partial_0 p^{\alpha}(t) =
\frac{2 g }{(2\pi)^2}\sum_{n\in\Z}
\sum_{\gamma=1}^N \left(1-\delta^{\alpha\gamma}\delta_{n,0}\right)
\frac{\hat \rho^{\alpha\gamma}(t,n) \hat \rho^{\gamma\alpha}(t,-n)}
{ \left(n +  g [q^{\alpha}(t)-q^{\gamma}(t) ]\right)^3}.
\end{equation}
Inserting Eq. (\ref{Aa}) and (\ref{A0_exp}) in (\ref{f3}) we get
the time evolution of $\hat{\rho}^{\alpha\beta}(t,n)$
\begin{eqnarray}
  \label{rhodot}
  \partial_0 \hat{\rho}^{\alpha\beta}(t,n) &=& -
-\frac{ig}{2\pi}\hat{\rho}^{\alpha\beta}(t,n)\left(a^\alpha(t)
-a^\beta(t)\right) - 
\nonumber \\ \nopagebreak
&&- \frac{ig}{2\pi}\sum_{k\in \Z}
  \sum_{\gamma=1}^N \left[ \left(1-\delta^{\alpha\gamma}\delta_{k,0}\right)
    \frac{\hat{\rho}^{\alpha\gamma}(t,k)
      \hat{\rho}^{\gamma\beta}(t,n-k)}{(k + g[q^\alpha(t)-q^\gamma(t)])^2}-    
 \right.\nonumber \\ \nopagebreak
&&- \left. \left(1-\delta^{\gamma\beta}\delta_{k,0}\right)
  \frac{\hat{\rho}^{\alpha\gamma}(t,n-k)
      \hat{\rho}^{\gamma\beta}(t,k)}{(k +
      g[q^\gamma(t)-q^\beta(t)])^2}\right]
\end{eqnarray}
We now have obtained the time evolution equations for our dynamical
system in a direct way. This result justifies our derivation of
the Hamiltonian (\ref{pq_hamiltonian}) in the last section:
{\em The equations (\ref{qdot})--(\ref{rhodot}) are identical with the
Hamiltonian equations obtained from the Hamiltonian
(\ref{pq_hamiltonian}) and the Poisson brackets
(\ref{canonical})--(\ref{rhohat_poi}).} (The proof of this
is a straightforward computation which we skip.)

It is worth to note that $p^\alpha$ and $q^\alpha$ are not necessarily
real. Denoting the real and imaginary parts of $p^\alpha$ by
$p^\alpha_r$ and $p^\alpha_i$, respectively, and similarly for
$q^\alpha$, we obtain from Eq. (\ref{qdot})
$$
 \partial_0 q_r^{\alpha}(t) = p_r^{\alpha}(t)\label{c1},
$$
and from Eq. (\ref{Pdot})
$$
\partial_0 p_r^{\alpha}(t) = {\rm Re} \left[ \frac{2 g }{(2\pi)^2}
\sum_{n\in\Z}\sum_{\gamma=1 }^N (1- \delta^{\alpha\gamma}\delta_{n,0})
\frac{\hat \rho^{\alpha\gamma}(t,n) \hat \rho^{\gamma\alpha}(t,-n)}
{ \left(n +  g [q_r^{\alpha}(t)- q_r^{\gamma}(t) + i [q_i^{\alpha}(t)
  -  q_i^{\gamma}(t) ] ) ]\right)^3} \right] \label{c3}
$$
and similarly for the imaginary parts.
Similarly, the time evolution of Eq. (\ref{rhodot}) can be split
in real and imaginary parts. In this general case, we can interpret 
the model as describing $2N$ particles
interacting with each other and spin degrees of freedom.

\subsubsection*{Weyl gauge}
We now discuss our dynamical system in the
Weyl gauge Eq. (\ref{weyl}). Imposing this condition,
(\ref{1})--(\ref{4}) are equivalent to
\begin{eqnarray}
\partial_0 A_1 = E,  \quad
\partial_0 E  = 0, \quad
\partial_0 \rho = 0 \nonumber \\ \nopagebreak
\partial_1 E + ig  [A_1,E] = \rho .
\label{lin}
\end{eqnarray}
which can be solved trivially (see Section \ref{sec_intsys}). We also
observe that {\em these equations are equivalent to the Hamiltonian
equations obtained from the Hamiltonian (\ref{weyl_hamiltonian}) and
the Poisson brackets (\ref{canon_rel}) and (\ref{rho_poi}).} (Again
the proof is a simple computation which we skip.)  We thus have
rigorously justified all results which in Section \ref{canon} using
the Hamiltonian method.

\section{Solvable non-linear systems}
\label{sec_intsys}
In this section we discuss how to obtain and solve dynamical systems 
which are special cases of the system defined by Eqs.\ 
(\ref{qdot})--(\ref{rhodot}).  We will impose additional constraints 
on the charges $\rho$ in order to obtain simpler systems which have 
natural physical interpretations and which can be solved explicitly.  
This will be done according to the following recipe:

\begin{enumerate}
\item Make an ansatz for the charge $\rho(t,x)$ which is parameterized
by a finite number of $x$-independent functions and which is
consistent with Eqs.\ (\ref{rho_cons}) and (\ref{rhodot}).
\item Insert this ansatz for $\rho$ in
the Eqs. (\ref{qdot})--(\ref{rhodot}). To obtain the Hamiltonian and
Poisson brackets giving rise to these equations, insert this ansatz
for $\rho$ in the Hamiltonian (\ref{pq_hamiltonian}) and the Poisson
bracket (\ref{rhohat_poi}).
\end{enumerate}

We will also show how to obtain a one-parameter family of Lax pairs
and conservation laws in our formalism.

We are interested in the solution of the system of differential
equations (\ref{qdot})--(\ref{rhodot}) with the initial conditions,
\begin{equation}  
q^\alpha(0)=q^\alpha_0,\quad p^\alpha(0)=p^\alpha_0, \quad
\hat\rho^{\alpha\beta}(0,n) = \hat\rho^{\alpha\beta}_0(n) .
\label{IC_DCG}
\end{equation} 
Our method of solving this initial value problem is based on
the results obtained in the last section. It was shown there that the
Eqs. (\ref{qdot})--(\ref{rhodot}) are obtained from a Yang-Mills gauge
theory by imposing the diagonal Coulomb gauge Eq.\ (\ref{DCG}).
Moreover, this very gauge theory in the Weyl gauge Eq.\ (\ref{weyl})
leads to the {\em linear} differential equations (\ref{lin}) which can
be solved trivially,
\begin{eqnarray}
  \label{A0}
  E(t,x)&=& E(0,x) , \nonumber \\ \nopagebreak
  A_1(t,x) &=& A_1(0,x) + E(0,x) t , \nonumber \\ \nopagebreak
  \rho(t,x) &=& \rho(0,x) 
\end{eqnarray}
with the initial data $E(0,x)$, $A_1(0,x)$ and $\rho(0,x)$ satisfying
Gauss' law Eq.\ (\ref{4}). It is important to note that our solution
(\ref{A0}) satisfies the Gauss' law for all $t$ if it satisfies it for
$t=0$. {}From our discussion in the last section it is obvious that
initial conditions for the Yang-Mills fields corresponding to the 
initial conditions Eq.\ (\ref{IC_DCG}) are, 
\begin{equation}  
A_1^{\alpha\beta}(0,x)=\delta^{\alpha\beta}q^\alpha_0,\quad
\int_{-\pi}^\pi dx E^{\alpha\alpha}(0,x) = 2\pi p^\alpha_0, \quad
\hat\rho^{\alpha\beta}(0,n) = \hat\rho^{\alpha\beta}_0(n) .
\label{IC_Weyl}
\end{equation}
Gauss' law Eq.\ (\ref{4}) then completely determines
$E^{\alpha\beta}(0,x)$ and thus the solution of our gauge theory
model in the Weyl gauge. In Section \ref{gauges}, we
also discussed how to construct the gauge transformation
transforming to the diagonal Coulomb gauge.
Combining these results we obtain the
following recipe for solving a large class of
non-linear integrable systems:

\begin{enumerate}
 \item Use the initial conditions Eq.\ (\ref{IC_Weyl}) to calculate
 $E(0,x)$ from the Gauss' law Eq.\ (\ref{4}). \item Take $A_1(t,x)$
 given by Eq. (\ref{A0}) and calculate $S(t,\pi)$ from
 Eq. (\ref{S_def}). \item The positions $q^\alpha(t)$ at time $t$ are
 given by the eigenvalues $\lambda^\alpha(t)$ of $S(t,\pi)$ according
 to $\lambda^\alpha(t)=e^{-ig 2\pi q^\alpha (t) }$, provided that
 $\lambda^\alpha(t')\neq \lambda^\beta(t')$ for all $\alpha\neq\beta$
 and $0\leq t'\leq t$. If this latter condition holds we say that
 {\em no particle collisions occur in the time interval $[0,t]$}.
 \item The solutions $\rho(t,x)$ are given by
 $U^{-1}(t,x)\rho(0,x)U(t,x)$ where $U(t,x)$ is the gauge
 transformation defined in Eq.\ (\ref{U}).
\end{enumerate}

We note that the no-collision condition above is due to the regularity
assumption which we needed to prove the existence of the diagonal
Coulomb gauge in Section \ref{gauges}.

We now show how to obtain Lax equations and conservation laws in our
formalism. For that we observe that one of our Yang-Mills equations,
$\partial_0 E(t,x) + ig[A_0(t,x),E(t,x)]=0$, has the form of a
Lax equation for arbitrary fixed $x$: For any $x$, $L(t)\equiv E(t,x)$ and
$M(t)\equiv igA_0(t,x)$ is a Lax pair, $\partial L/\partial t +[M,L]=0$. We
thus have a {\em one-parameter family of Lax pairs}. By a standard
argument this implies that $tr[E(t,x)^n]$ is invariant under time
evolution, for arbitrary positive integer $n$ and $x\in [-\pi,\pi]$.
We have already found these conservation laws by a different argument
at the end of Section \ref{canon}.

In case the computation here can be done explicitly we obtain
explicit solutions, Lax pairs and conservation laws. We will now
discuss several examples of increasing complexity where this is
possible. Before that we state three identities which will be
important for us.\\

\vspace*{0.5cm} \noindent {\bf Three identities.} The first identity
we will need is well-known, 
\begin{equation}  
\label{id0}
\sum_{n\in {\bf Z}} \frac{1}{(n+r)^2} =
\frac{\pi^2}{\sin^2{(\pi r)}} ,\quad r\in\C
\end{equation}
(see e.g.\ \cite{Grad}). The second is a generalization of the first
\begin{equation}
\label{id}
\sum_{n\in\Z}\frac{e^{ins} }{(n+r)^2} =
e^{-i rs_{2\pi} }\left(\frac{\pi^2}{\sin^2(\pi r)}
+ i \pi s_{2\pi} \cot(\pi r) -\pi|s_{2\pi}|
\right) , \quad r\in\C,s\in\R ,
\end{equation}
where
\begin{equation}  
s_{2\pi} : = s - 2\pi n\quad \mbox{with $n\in\Z$ such that
$-\pi\leq s - 2\pi n <\pi$}.
\end{equation}
Finally,
\begin{equation}
\label{id1}
\sum_{n\neq 0}\frac{e^{ins} }{n^2} = \frac{(s_{2\pi})^2}{2}-\pi|s_{2\pi}| +
\frac{\pi^2}{3}, \quad s\in\R .
\end{equation}
Proofs of Eqs.\ (\ref{id}) and (\ref{id1}) are given in Appendix
\ref{derivation}.

\subsection{Example 1: Spin CM model with $v(r)=r^{-2}$}
\label{sec_simpex}
As the simplest example we now show how to obtain and solve the
Calogero model and certain spin generalizations thereof.

For that we make the following ansatz,
\begin{equation}
  \hat{\rho}^{\alpha\beta}(t,n) =
  \delta_{n,0}(1-\delta^{\alpha\beta})2\pi g \S^{\alpha\beta}(t) .
  \label{rho1}
\end{equation}
It is important to note that Eq.\ (\ref{rhodot}) implies that
$\partial_0\hat{\rho}^{\alpha\beta}(t,n)=0$ for $n\neq 0$ and
$\partial_0\hat{\rho}^{\alpha\alpha}(t,0)=0$. Hence the
ansatz Eq.\ (\ref{rho1}) is consistent, and the functions
$\S^{\alpha\beta}(t)$ for $\alpha\neq\beta$, together with the
$p^\alpha(t)$ and $q^\alpha(t)$, are the dynamical
variables of a dynamical system.

{}From the Hamiltonian
(\ref{pq_hamiltonian}) and the Poisson brackets
(\ref{canonical})--(\ref{rhohat_poi}) we obtain
\begin{eqnarray}
  {\cal H} &=& \frac{1}{2}\sum_{\alpha=1}^N (p^\alpha)^2 +
  \frac{1}{2}\sum_{\stackrel{\alpha,\beta = 1}{\alpha\neq\beta}}^N
  \frac{\S^{\alpha\beta}\S^{\beta\alpha}}{(q^\alpha-q^\beta)^2}\label{rham}\\
  \{p^\alpha,q^\beta\} &=& \delta^{\alpha\beta}\label{rpoi1}\\
  \{\S^{\alpha\beta} , \S^{\alpha'\beta'}\} &=&
  i ( \S^{\alpha\beta'}\delta^{\beta\alpha'} -
  \S^{\alpha'\beta}\delta^{\beta'\alpha})\label{rpoi2}
\end{eqnarray}
which determines the equations of motion for the dynamical variables.
Moreover, the initial conditions are
\begin{equation}  
q^\alpha(0)=q^\alpha_0, \quad p^\alpha(0)=p^\alpha_0, \mbox{ and }
\S^{\alpha\beta}(0) = \S^{\alpha\beta}_0 \quad
\mbox{ all real.}
\end{equation}

We now have to determine the corresponding initial conditions for the
Yang-Mills fields. Using
$E^{\alpha\beta}(0,x)=\frac{1}{2\pi}\sum_{n\in\Z}
\hat{E}^{\alpha\beta}(0,n)e^{inx}$ and Eqs.\ (\ref{p_alpha}) and
(\ref{rhoexp}) we obtain 
\begin{equation}  
 E^{\alpha\beta}(0,x) =
p^{\alpha}_0\delta^{\alpha\beta} + (1-\delta^{\alpha\beta})
\frac{\S^{\alpha\beta}_0}{i[q^\alpha_0-q^\beta_0]} .
\label{rE}
\end{equation}
Moreover, we recall
$A_1^{\alpha\beta}(0,x)=\delta^{\alpha\beta}q^{\alpha}_0$.
Inserting this and Eq.\ (\ref{rE}) in Eq.\ (\ref{A0}) we find
that $A_1(t,x)=A_1(t)$ is independent of $x$,
\begin{equation}
  A_{1}^{\alpha\beta}(t) = q^{\alpha}_0\delta^{\alpha\beta} +
  \left(p^{\alpha}_0\delta^{\alpha\beta} + (1-\delta^{\alpha\beta})
    \frac{\S^{\alpha\beta}_0}{i[q^\alpha_0-q^\beta_0]}
  \right)t.
\label{explicit}
\end{equation}
This is the solution of the Yang-Mills equations. To find the solution
for our dynamical system we need to transform this to the diagonal Coulomb
gauge Eq.\ (\ref{DCG}). This is simple here: all we need to do is
diagonalize the matrix $A_1(t)$. Furthermore, our discussion in
Section \ref{sec_intsys} implies that $\S^{\alpha\beta}(t)$ can be obtained
from the matrix $U(t)$ diagonalizing
$A_{1}(t)$. We thus obtain the following

\vspace*{0.3cm} 
\noindent{\bf Result 1:} {\em Provided that no particle collisions
occur in the time interval $[0,t]$, the solution $q^\alpha(t)$ of the
initial value problem corresponding to the Hamiltonian (\ref{rham})
and the Poisson brackets (\ref{rpoi1}) and (\ref{rpoi2}) is given by
the eigenvalues of the matrix $A_{1}(t)$ in Eq.\ (\ref{explicit}),
$$
U(t)^{-1}A_1(t)U(t)={\rm diag}\left( q^1(t), q^2(t), \ldots,
q^N(t) \right) .
$$
Moreover, the diagonalizing matrix $U(t)$ gives the time evolution of
the spin degrees of freedom, $\S(t)=U(t)^{-1}\S(0)U(t)$.}

\vspace*{0.3cm} 
We thus have recovered the integrable systems and their solution
previously found in Ref.\ \cite{Woj}.
In the present case we only get one Lax pair (the Lax pair family
is independent of $x$) which coincides with the one given in
Ref.\ \cite{Woj}.

Note that the special case where all initial
$\S^{\alpha\beta}$ are identical, $\S^{\alpha\beta}(0) = e$,
corresponds to the Calogero model, i.e., the Hamiltonian Eq.\ (\ref{CMS})
with $v(r)=r^{-2}$. To see this,
it is important to notice that the functions $a^\alpha(t)$ in Eq.\
(\ref{rhodot}) can be chosen such that $\S^{\alpha\beta}(t) = e$ for all $t$.
Indeed, with
\begin{equation}
  a^\alpha(t) = -\frac{2\pi e}{g}
  \sum_{\stackrel{\gamma=1}{\gamma\neq\alpha}}^N
  \frac{1}{(q^\alpha(t)-q^\gamma(t))^2}
  \label{rresid}
\end{equation}
we obtain $\partial_0\S^{\alpha\beta}(t)=0$. 

\vspace*{0.3cm}\noindent
{\em Remark:} It would be interesting to find and study a case
where collisions occur.

\subsection{Example 2: CM model with $v(r) = a^2\sin^{-2}(ar)$}
\label{sutherland}
In this section we show in detail how to recover the known solution of
the Sutherland model, i.e.,\ the Hamiltonian Eq.\ (\ref{CMS}) with the
potential $v(r) = a^2\sin^{-2}(ar)$. This is a simple special case
and warm-up for what is discussed in Section \ref{general}.

To obtain the Hamiltonian of the Sutherland model we make the ansatz
\begin{equation}  
  \hat{\rho}^{\alpha\beta}(0,n) = (1-\delta^{\alpha\beta})2\pi
  ge\quad
  \forall n\in \Z
  \label{srho}
\end{equation}
equivalent to
\begin{equation}
                \label{ans1}
   \rho^{\alpha\beta}(0,x) =
   (1-\delta^{\alpha\beta}) 2 \pi g e \delta(x).
   \label{srhox}
\end{equation}
Choosing
\begin{equation}
  a^\alpha(t) = \sum_{k\in \Z} \sum_{\gamma=1}^N
  (1-\delta^{\alpha\gamma}\delta_{k,0}) \frac{-2\pi
  ge}{(k+g[q^\alpha(t)-q^\gamma(t)])^2}
  \label{sresid}
\end{equation}
implies $\partial_0\hat{\rho}^{\alpha\beta}(t,n)=0$, as can be seen by
inserting Eq. (\ref{sresid}) in Eq. (\ref{rhodot}). We thus obtain
$\hat{\rho}^{\alpha\beta}(t,n) = (1-\delta^{\alpha\beta})2 \pi g e$
for all $t$, and $p^\alpha$ and $q^\alpha$ are in fact the only
dynamical variables in the system.

Inserting Eq.\ (\ref{srho}) into the Hamiltonian Eq.\
(\ref{pq_hamiltonian}) and using Eq.\ (\ref{id0}) we obtain
\begin{equation}
  {\cal H} = \frac{1}{2}\sum_{\alpha=1}^N (p^\alpha)^2 +
  \frac{e^2}{2}\sum_{\stackrel{\alpha,\beta = 1}{\alpha\neq\beta}}^N
  \frac{(\pi g)^2}{\sin^2(\pi g[q^\alpha-q^\beta])} \label{sham}
  \label{spoi}
\end{equation}
which for $g=a/\pi$ equals the Hamiltonian of the Sutherland model as
discussed in Section \ref{intro}.\\

We now show how to solve the initial value problem for the Hamiltonian
equations following from the Hamiltonian Eq.\ (\ref{spoi}) and the
Poisson brackets $\{q^\alpha,p^\beta\}= \delta^{\alpha\beta}$. We
first have to determine the initial conditions for the Yang-Mills
fields corresponding to $q^\alpha(0)=q^\alpha_0$ and
$p^\alpha(0)=p^\alpha_0$ all real. For $t=0$ we can write Gauss' law
Eq.\ (\ref{4}) as
\begin{equation}
  \label{s22}
  \partial_1 \left( e^{ig Q_0 x} E(0,x) e^{-i
        gQ_0x} \right) = e^{ig Q_0x}\rho(0,x)
  e^{-i g Q_0x}
\end{equation}
where $Q_0=Q(0)={\rm diag}(q_0^1,\ldots,q_0^N)$ and
$\rho(0,x)$ is given by Eq.\ (\ref{srhox}).
Since $\rho(0,x)=0$ except for $x=0$, it follows that $e^{ig Q_0
    x} E(0,x) e^{-ig Q_0x}$ is constant in the intervals
$-\pi\leq x < 0$ and $0 < x <\pi$. We therefore can write
\begin{equation}
  E(0,x) = e^{-ig Q_0 x}B_\pm e^{ig Q_0 x}\quad \mbox{ for
  $x\gtlt 0$}
  \label{sinite}
\end{equation}
with $B_\pm$ constant matrices.

To determine the matrices $B_\pm$ we integrate Eq.\ (\ref{s22}) from
$0-\epsilon$ to $0+\epsilon$ and then take the limit
$\epsilon\downarrow 0$. This gives
$$
  B_{+}^{\alpha \beta} - B_{-}^{\alpha\beta} =
  (1-\delta^{\alpha\beta})2\pi ge .
$$
We also recall $\int_{-\pi}^\pi dx E^{\alpha\alpha}(0,x) =
2\pi p^\alpha_0$ which yields
$$
\frac{1}{2}
\left(B_-^{\alpha\alpha} + B_+^{\alpha\alpha}\right) =  p^\alpha_0 .
\label{bpm3}
$$
Moreover, the boundary condition $E(0,-\pi)=E(0,\pi)$ implies 
$e^{ 2i\pi g Q_0}B_- e^{-2i\pi g Q_0} = B_+$, or equivalently
$$
B_-^{\alpha\beta} e^{2i\pi g(q_0^\alpha-q_0^\beta)} = B_+^{\alpha\beta}.
$$
A straightforward computation shows that
these equations determine the elements of
the matrices $B_-$ and $B_+$ as follows,
\begin{equation}
  \label{sB}
  B_\pm^{\alpha\beta} = \delta^{\alpha\beta} p^{\alpha}_0 +
  \left(1-\delta^{\alpha\beta}\right) \frac{e\pi g
    e^{\pm i\pi g[q^{\alpha}_0-q^{\beta}_0]}  }
  {i\sin(\pi g[q^{\alpha}_0-q^{\beta}_0] )} .
\end{equation}
Inserting the initial condition $A_1(0,x)=Q_0$
and Eq.\ (\ref{sinite}) in Eq.\ (\ref{A0})
we finally obtain the explicit
solution of the Yang-Mills equation in
the Weyl-gauge,
\begin{equation}
A_1(t,x) = e^{-ig Q_0 x} \left( Q_0 + B_\pm t\right) e^{ig Q_0 x}
\quad \mbox{ for  $x\gtlt 0$}
\label{A1S}
\end{equation}
(we used $Q_0=e^{-ig Q_0 x}Q_0 e^{ig Q_0 x}$).

We now determine the gauge transformation which transforms this
solution to the diagonal Coulomb gauge. For that we
need to solve Eq. (\ref{S_def}) for $S(t,x)$. We define
$\tilde S(t,x) = e^{ig Q_0 x} S(t,x)$
and observe that Eqs.\ (\ref{S_def}) and (\ref{A1S}) imply
\begin{equation}
  \partial_1 \tilde S(t,x) + i g B_\pm t\tilde S(t,x) =0
  \quad \mbox{ for
  $x\gtlt 0$} .
  \label{stileq}
\end{equation}
This latter equation can be solved trivially,
\begin{equation}
  \tilde{S}(t,x) = e^{-ig B_\pm tx }\tilde{S}(t,0)\quad \mbox{ for
  $x\gtlt 0$} .
\end{equation}
We thus obtain $\tilde S(t,\pi)= e^{-i \pi g B_+t}\tilde S(t,0)$ and $\tilde
S(t,0)= e^{-i \pi g B_- t}\tilde S(t,-\pi)$, and
\begin{eqnarray*}
  S(t,\pi) &=& e^{-i\pi g Q_0 } \tilde{S}(t,\pi) = e^{-i\pi g Q_0}
  e^{-i\pi gB_+t} \tilde{S}(t,0) =  \\
  &=& e^{-i\pi g Q_0 } e^{-i\pi gB_+t} e^{-i\pi gB_-t} \tilde{S}(t,-\pi)=  \\
  &=& e^{-i\pi g Q_0} e^{-i\pi gB_+t} e^{-i\pi gB_-t} e^{-i\pi g Q_0}
\end{eqnarray*}
where we used $S(t,-\pi)=I$.

In the present case we can easily prove that the Yang-Mills field
$A_1(t,x)$ Eq.\ (\ref{A1S}) is generic, i.e., if $q^\alpha_0\neq
q^\beta_0$ for all $\alpha\neq\beta$ then particle collisions never
occur. This follows from the fact that the Hamiltonian ${\cal H}$
in Eq. (\ref{spoi}) is conserved under the time evolution, and this
implies an upper bound on $\sin^{-2}(\pi g[q^\alpha-q^\beta])$. There
exists therefore an $\eps>0$ such that $|q^{\alpha}(t)-
q^{\beta}(t)-n/g|>\eps$ for all $\alpha\neq\beta$, $t\in\R$, and all
integers $n$. Our discussion in Section \ref{gauges} implies that the
latter is equivalent to $A_1(t,x)$ being generic, i.e., regular at all
times $t$.

{}From the discussion in Section \ref{sec_intsys} we thus obtain the

\vspace*{0.3cm} 
\noindent{\bf Result 2:}
{\em The solution
of the initial value problem for the Sutherland model Eq.\ (\ref{sham})
is given by the eigenvalues $\lambda^\alpha(t)$ of the matrix
\begin{equation}  
\label{ss1}
e^{-i\pi g Q_0} e^{-i\pi gB_+t} e^{-i\pi gB_-t} e^{-i\pi g Q_0} 
\end{equation}
with the matrix elements of $B_\pm$ given in Eq. (\ref{sB}) and
$Q_0={\rm diag}(q_0^1,\ldots,q_0^N)$, according to\footnote{the
branch of the log is fixed by continuity of $q^\alpha(t)$}
$q^\alpha(t)=-\log (\lambda^\alpha(t) )/(2\pi ig)$.}

\vspace*{0.3cm}
\noindent{\em Remark:}
To compare this with the well-known result reviewed in \cite{OPcl}
we observe that
$
B_\pm = e^{\pm i\pi g Q_0 } B e^{\mp i\pi g Q_0},
$
where $$B^{\alpha\beta} = \delta^{\alpha\beta} p^{\alpha}_0 +
   \left(1-\delta^{\alpha\beta}\right) \frac{e\pi g}
{i\sin(\pi g[q^{\alpha}_0-q^{\beta}_0] )}. $$
Using this it is easy to
see that the matrix in Eq.\
(\ref{ss1}) has the same eigenvalues as the matrix
\begin{equation}  
e^{-i\pi g Q_0}e^{-2i\pi gBt}e^{-i\pi g Q_0},
\end{equation}
which shows that the result above is identical with the one derived
in Ref.\ \cite{OPcl}.
\vspace*{0.3cm}

We finally show how to recover the Lax pair and conservation laws for
the Sutherland model in our formalism. We observe that our
computation of $E(t,x)$ at $t=0$ in terms of
$q^\alpha_0=q^\alpha(t=0)$ and $p^\alpha_0=p^\alpha(t=0)$ immediately
generalizes to $t\neq 0$. From our solution at $t=0$ we therefore can
read off that 
\begin{equation}  
 E(t,x) = e^{-ig(x\mp \pi)Q(t)}B(t)e^{ig(x\mp
\pi)Q(t)} \quad \mbox{ for $x\gtlt 0$} 
\end{equation} 
with 
\begin{equation}    
B^{\alpha\beta}(t) = \delta^{\alpha\beta} p^{\alpha}(t) +
\left(1-\delta^{\alpha\beta}\right) \frac{e\pi g} {i\sin(\pi
g[q^{\alpha}(t)-q^{\beta}(t)] )} . 
\end{equation} 
Inserting
Eqs.\ (\ref{srho}) and (\ref{sresid}) in Eq.\ (\ref{A0_exp}) and using
$A_0(t,x)=\frac{1}{2\pi}\sum_{n\in\Z}e^{i nx}
\hat{A}_0(t,n)$
together with the identities in Eqs.\ (\ref{id}) and (\ref{id1}), we obtain
after a simple computation
\begin{eqnarray*}
  A_0^{\alpha\beta}(t,x)&=&-\delta^{\alpha\beta}eg\left(
\sum_{\stackrel{\gamma=1}
   {\gamma\neq\alpha}}^N \frac{\pi^2}{\sin^2(\pi g
   [q^\alpha(t)-q^\gamma(t)] )} +\frac{\pi^2}{3} \right) \\
   &+&(1-\delta^{\alpha\beta})eg e^{-igx_{2\pi}[q^\alpha(t)-q^\beta(t)]}
   \times\\
   &\times&
   \left( \frac{\pi^2}{\sin^2(\pi g [q^\alpha(t)-q^\beta(t)])} + i\pi x_{2\pi}
   \cot(\pi g[q^\alpha(t)-q^\beta(t)]) - \pi |x_{2\pi}| \right).
   \label{lax2}
\end{eqnarray*}
We thus have found a one-parameter family of Lax pairs for the
Sutherland model. However, due to cyclicity of trace, the
corresponding conservation laws are independent of $x$:
$tr[E(t,x)^n]=tr[B(t)^n]$. It is interesting to note that the
standard Lax pair for the Sutherland model, derived e.g.\ in Ref.\
\cite{OPcl}, is recovered from our Lax pair family above by setting
$x=\pi$.

\subsection{Example 3: CM model with $v(r) = a^2\sinh^{-2}(ar)$}
\label{suthsh}
We now make a simple but important observation:
In Example 2 above we can allow for general complex initial
data $q^\alpha_0$ and $p^\alpha_0$ and thus obtain time evolution
equations for complex valued functions $q^\alpha(t)$ and
$p^\alpha(t)$. In this case all our discussion in Example 2 above
goes through,
except the argument showing that no particle collisions occur.
Thus our Result 2 above holds also for the
general complex case at least locally in time, i.e., until
the first particle collision occurs.

An important special case is obtained if $e$ is real and all
$q^\alpha_0$ and $p^\alpha_0$ are chosen purely imaginary, i.e., they
are real if multiplied with $i$. It is easy to convince oneself
that the results for this case we can obtain from the results in
Example 2 simply by the following replacements 
\begin{equation}  
\label{to1}
q^\alpha(t)\to iq^\alpha(t),\quad p^\alpha(t)\to i p^\alpha(t)
\end{equation}
where $q^\alpha(t)$ and $p^\alpha(t)$ real. The Hamiltonian and
Poisson brackets then get minus signs which can be conveniently
removed by the transformation,
\begin{equation}  
\label{to2}
{\cal H}\to
-{\cal H}, \quad \{\cdot,\cdot\}\to - \{\cdot,\cdot\} ,
\end{equation} 
which
leaves the Hamilton equations $\partial_0 X=\{{\cal H},X\} $ invariant.
The resulting initial value
problem thus is equivalent to the one obtained from the Hamiltonian
\begin{eqnarray}
  {\cal H} = \frac{1}{2}\sum_{\alpha=1}^N (p^\alpha)^2 +
  \frac{e^2}{2}\sum_{\stackrel{\alpha,\beta = 1}{\alpha\neq\beta}}^N
  \frac{(\pi g)^2}{\sinh^2(\pi g[q^\alpha-q^\beta])} \label{sham1}
\end{eqnarray}
and the standard Poisson brackets $\{q^\alpha,p^\beta\}=
\delta^{\alpha\beta}$. This initial value problem is identical with
the one for the Sutherland model with the potential $v(r) =
a^2\sinh^{-2}(ar)$ and $a=\pi g$ \cite{OPcl}. In this case we can
argue as in Example 2 that no particle collisions can occur, i.e.,
$|q^\alpha(t)- q^\beta(t)|>\eps$ for some $\eps>0$ and all $t\in\R$
and all $\alpha\neq\beta$. Result 2 above then implies:

\vspace*{0.3cm} 
\noindent{\bf Result 3:} {\em The solution of the initial value
problem for the Sutherland model defined by the Hamiltonian in Eq.\
(\ref{sham1}) is given by the eigenvalues $\lambda^\alpha(t)$ of the
following matrix, 
\begin{equation}  
 e^{\pi g Q_0} e^{\pi gB_+t} e^{\pi gB_-t} e^{\pi
g Q_0} 
\end{equation} 
with 
\begin{equation}  
 B_\pm^{\alpha\beta} = \delta^{\alpha\beta}
p^{\alpha}_0 - \left(1-\delta^{\alpha\beta}\right) \frac{e\pi g
e^{\mp\pi g[q^{\alpha}_0-q^{\beta}_0]} } {\sinh(\pi
g[q^{\alpha}_0-q^{\beta}_0] )} 
\end{equation} 
and $Q_0={\rm
diag}(q_0^1,\ldots,q_0^N)$, according to $q^\alpha(t)=
\log(\lambda^\alpha(t))/(2\pi g)$. }

\vspace*{0.3cm}
 
As in Example 2 one obtain Lax pair etc.\ and can check that
what we obtain is equivalent to the known result for this
model given in \cite{OPcl}.

\subsection{Example 4: Generalizing the Sutherland models}
\label{general}
We now discuss a general case for a spin-particle dynamical system
equivalent to a Yang-Mills gauge theory as discussed in Section
\ref{sec_YM} and which we can solve explicitly. Our result here
generalizes the one in Ref.\ \cite{BL}.

We choose the charge distribution of the form
\begin{equation}
\label{rho}
\rho^{\alpha\beta}(t,x) \equiv \sum_{j=1}^m
\rho^{\alpha\beta}_{j}(t)\delta(x-x_j)
\end{equation}
where $m$ is some positive integer and we assume, without loss of
generality, 
\begin{equation}  
x_0\equiv -\pi < x_1 < x_2< \ldots < x_m<x_{m+1}\equiv
\pi . 
\end{equation} 
This charge distribution describes matter charges
localized at the points $x_j$, $1\leq j\leq m$. Using Eq. (\ref{3})
it is easy to see that this ansatz is preserved under time evolution.
Moreover, using
\begin{equation}
\label{rhohat}
\hat\rho^{\alpha\beta}(t,n)=\sum_{k=1}^m \rho_k^{\alpha\beta}(t)
e^{-in x_k}
\end{equation}
we see that Eq.\ (\ref{rho_cons}) implies  the constraint
\begin{equation}
\label{condition}
\sum_{j=1}^m \rho_j^{\alpha\alpha}(0) = 0
\quad \forall \alpha=1,\ldots,N
\end{equation}
on possible initial conditions for the $\rho^{\alpha\beta}_{j}$.

Using the identities (\ref{id}) and (\ref{id1}),
the Hamiltonian in Eq.\ (\ref{pq_hamiltonian}) becomes (we specialize
to $a^\alpha(t)=0$ for simplicity)
\begin{equation}
  \label{nov_ham}
  {\cal H}=\frac{1}{2}\sum_{\alpha=1}^N  (p^\alpha)^2 + \frac{1}{2}
  \sum_{ \stackrel{\alpha,\beta=1}{\alpha\neq \beta} }^N
  \sum_{j,k=1}^m v_{jk} (q^\alpha-q^\beta) \rho_k^{\alpha\beta}
  \rho_j^{\beta\alpha} + \frac{1}{2} \sum_{ \alpha=1}^N \sum_{j,k=1}^m
  c_{jk} \rho_k^{\alpha\alpha} \rho_j^{\alpha\alpha},
\end{equation}
where
\begin{equation}
\label{V}
v_{jk} (r) = \frac{1}{4}e^{-i g r x_{jk} }
\left(\frac{1}{\sin^2(\pi g r)} + \frac{i x_{jk}}{\pi} \cot(\pi g r)
-\frac{|x_{jk}|}{\pi}\right)
\end{equation}
and
\begin{equation}
\label{cjk}
c_{jk} = \frac{x_{jk}^2}{8\pi^2} - \frac{|x_{jk}|}{4\pi} +
\frac{1}{12}.
\end{equation}
Here $x_{jk}$ is defined as
\begin{equation}
x_{jk} = (x_j-x_k)_{2\pi}.\nonumber
\end{equation}
The Poisson brackets for this system are:
\begin{eqnarray}
  \label{pb1}
  &\{p^\alpha,q^\beta\}& = \delta^{\alpha\beta}\\
  \label{pb2}
  &\{\rho_j^{\alpha\beta},\rho_k^{\alpha'\beta'} \}& =
  ig 2\pi \delta_{jk}\left(
    \delta^{\beta\alpha'}\rho_j^{\alpha\beta'} -
    \delta^{\beta'\alpha}\rho_j^{\alpha'\beta} \right).
\end{eqnarray}

We now solve the equations of motion for the system given by the
Hamiltonian (\ref{nov_ham}) and the Poisson brackets
(\ref{pb1})--(\ref{pb2}). More general than in \cite{BL}, we do not
assume that $q^\alpha(t)$ etc.\ all are {\em real}.

We start with Gauss' law Eq.\ (\ref{s22})
where $Q_0=Q(0)$ is given by (\ref{DCG}) and $\rho(0,x)$ by
(\ref{rho}). Since $\rho(0,x)=0$ except for $x=x_j$, we obtain from that
$E(0,x)= e^{-ig Q_0 x}B_j e^{ig Q_0 x}$, when $x_j<x<x_{j+1}$,
$j=0,\ldots,m$, where $B_j$ is some constant matrix (we recall that
$x_0=-\pi$ and $x_{m+1}=\pi$). To determine the matrices
$B_j$, we integrate Eq.\ (\ref{s22}) from $x_j-\epsilon$ to
$x_j+\epsilon$ and then take the limit
$\epsilon\downarrow 0$. This gives the recursion relations $ B_{j} -
B_{j-1} = e^{ig Q_0x_j}\rho_j(0) e^{-ig Q_0 x_j}$, and the condition
$E(0,-\pi)=E(0,\pi)$ implies $e^{ 2ig Q_0\pi}B_0 e^{-2ig Q_0\pi } =
B_m$. Combining these equations and using the definition
Eq. (\ref{qdot}) of $p^\alpha$, we obtain after a straightforward
calculation,
\begin{eqnarray}
  \label{B}
  B_j^{\alpha\beta} &=& \delta^{\alpha\beta}\left( p^{\alpha}(0)
    +\sum_{\ell = 1}^j\rho^{\alpha\alpha}_\ell(0) -
    \sum_{i = 1}^m \frac{x_{i+1}-x_i}{2\pi}\sum_{\ell=1}^i
    \rho^{\alpha\alpha}_\ell(0)
  \right) +\nonumber \\ \nopagebreak &+& 
  \left(1-\delta^{\alpha\beta}\right) \sum_{\ell=1}^m \frac{
    \rho_\ell^{\alpha\beta}(0)
    e^{ ig [q^{\alpha}(0)-q^{\beta}(0)][x_\ell + \pi{\rm sgn}(x_j-x_\ell)] }  }
  {2i\sin(g\pi[q^{\alpha}(0)-q^{\beta}(0)] )}
\end{eqnarray}
with ${\rm sgn}(x)=1$ for $x\geq 0$ and $-1$ for $x<0$.
With that, the solution of the Yang-Mills equations in the Weyl gauge,
Eq.\ (\ref{A0}), gives
\begin{equation}
  A_1(t,x) = e^{-ig Q_0 x} \left( Q_0 + B_j t\right) e^{ig Q_0 x} \;
{\rm for} \; x_j<x<x_{j+1}.\nonumber
\end{equation}

Next, we solve $\partial_1 S(t,x)+ig A_1(t,x) S(t,x)=0$ which is
equivalent to
\begin{equation}
  \partial_1 \tilde S(t,x) + ig B_j t\tilde S(t,x)=0 \quad \mbox{
    for $x_j<x<x_{j+1}$}\nonumber
\end{equation}
for $\tilde S(t,x) = e^{ig Q_0 x}
S(t,x)$. This implies $\tilde S(t,x)= e^{-i g B_j
t(x-x_j)}\tilde S(t,x_j)$ for $x_j<x<x_{j+1}$, thus
\begin{eqnarray}
\label{Sj}
 S(t,x_{j+1}) &=& e^{-ig Q_0 x_{j+1} }
e^{-ig B_j t(x_{j+1}-x_j)}\tilde S(t,x_j )
= \ldots = \\ \nopagebreak
&=& e^{-ig Q_0 x_{j+1} }
e^{-ig B_j t(x_{j+1}-x_j) }e^{-ig B_{j-1} t (x_j-x_{j-1})}
\cdots
e^{-ig B_0 t(x_1+\pi)} e^{ -ig Q_0 \pi}, \nonumber
\end{eqnarray}
where we used $S(t,-\pi)=I$. Especially (for $j=m$),
\begin{eqnarray}
\label{W}
S(t,\pi) = e^{-ig Q_0 \pi}
e^{-ig B_m t(\pi-x_m)} \cdots \nonumber \\ \nopagebreak \times
e^{-ig B_{m-1} t(x_{m}-x_{m-1})}\cdots
e^{-ig B_{0} t(x_{1}+\pi)}
e^{ -ig Q_0 \pi}.
\end{eqnarray}
{}From our discussion in Section \ref{sec_intsys} we can obtain the
$q^\alpha(t)$ from the eigenvalues of this matrix. Moreover,
$$
  \rho_j(t)=U(t,x_j)^{-1}\rho_j(0)U(t,x_j),
$$
where $U(t,x_j)$ is given by (\ref{U}). We thus obtain the

\vspace*{0.3cm} 
\noindent{\bf Result 4:} {\em Provided no particle collisions occur in
the time interval $[0,t]$, the solution $q^\alpha(t)$ of the initial
value problem defined by the Hamiltonian (\ref{nov_ham}) and the
Poisson brackets (\ref{pb1})--(\ref{pb2}) is given by the eigenvalues
$\lambda^\alpha(t)$ of the matrix $S(t,\pi)$ defined in Eqs. (\ref{W})
and (\ref{B}), according to\footnote{the branch of the log is fixed by
continuity of $q^\alpha(t)$}
$q^\alpha(t)=-\log(\lambda^\alpha(t))/(2\pi ig)$. Moreover, the time
evolution of the spin degrees of freedom is given by
\begin{equation}
\rho_j(t)=e^{-ig(x_{j}+\pi)Q(t)}V(t)^{-1}S(t,x_j)^{-1}
\rho_j(0)S(t,x_j)V(t) e^{ig(x_{j}+\pi)Q(t)}
\end{equation}
where $S(t,x_j)$ is defined in Eqs.\ (\ref{Sj}) and (\ref{B}),
$Q(t)={\rm diag}(q_{1}(t),q_{2}(t),\ldots q_{N}(t)  ) $
and $V(t)$ is the matrix diagonalizing $S(t,\pi) $, i.e.\
$$
V(t)^{-1} S(t,\pi) V(t) = e^{-ig2\pi Q(t)} .
$$
}

\vspace*{0.3cm} 
As discussed in the beginning of this section, $tr[E(t,x)^n]$ is
conserved under time evolution. {}For the special choice of charges
studied above, we can evaluate $E(t,x)$ and obtain $E(t,x)= e^{-ig
Q(t) x}B_j(t)e^{ig Q(t) x}$ for $x_j<x<x_{j+1}$ where $Q(t)$ and
$B_j(t)$ are as in Eqs. (\ref{DCG}) and (\ref{B}) but with
$q^\alpha(0)$, $p^\alpha(0)$, and $\rho_j^{\alpha\beta}(0)$ replaced
by $q^\alpha(t)$, $p^\alpha(t)$, and $\rho_{j}^{\alpha\beta}(t)$,
i.e.,\ the solution of the initial value problem which we solved
above. Using cyclicity of the trace, we conclude that $tr[B_j(t)^n]
$, for an arbitrary positive integer $n$ and $j=1,\ldots m$, are time
independent: Each of them is a conservation law.

The one-parameter family of Lax equations is $\partial_0 E(t,x) +
ig[A_0(t,x),E(t,x)]=0$. We now compute the corresponding family of
Lax pairs explicitly, similarly as in Section {\ref{sutherland}}. We
obtain after a straightforward calculation,
\begin{eqnarray}
  E(t,x) &=& \delta^{\alpha\beta}\left( p^{\alpha}(t)
    +\sum_{\ell = 1}^j\rho^{\alpha\alpha}_\ell(t) -
    \sum_{i = 1}^m \frac{x_{i+1}-x_i}{2\pi}\sum_{\ell=1}^i
    \rho^{\alpha\alpha}_\ell(t)
  \right) +  \nonumber \\ \nopagebreak &+&  
  \left(1-\delta^{\alpha\beta}\right) e^{-igx[q^\alpha(t)-q^\beta(t)]}
  \sum_{\ell=1}^m \frac{
    \rho_\ell^{\alpha\beta}(t)
    e^{ ig [q^{\alpha}(t)-q^{\beta}(t)][x_\ell + \pi{\rm sgn}(x_j-x_\ell)] }  }
  {2i\sin(g\pi[q^{\alpha}(t)-q^{\beta}(t)] )}
\end{eqnarray}
and
\begin{eqnarray*}
   A_0^{\alpha\beta}(t,x)&=&-\delta^{\alpha\beta} \frac{1}{2\pi} \sum_{k=1}^m
   \rho_k^{\alpha\alpha}(t) \left( \frac{(x-x_k)_{2\pi}^2}{2}
   -\pi|(x-x_k)_{2\pi}| + \frac{\pi^3}{3} \right) + \\
   &+&  (1-\delta^{\alpha\beta}) \frac{\pi}{2} \sum_{k=1}^m
   \rho_k^{\alpha\beta}(t) e^{-ig(x-x_k)_{2\pi}[q^\alpha(t)-q^\beta(t)]}
   \left( \frac{1}{\sin^2(\pi g [q^\alpha(t)-q^\beta(t)])}  
   \right.  + \\ &+& \left. 
   \frac{i(x-x_k)_{2\pi}} {\pi}\cot(\pi g[q^\alpha(t)-q^\beta(t)]) -
   \frac{|(x-x_k)_{2\pi}|}{\pi} \right).
\end{eqnarray*}

Note that for $m=1$, the dynamics of the spin and the particles
decouple, and we recover the results in Sections \ref{sutherland} and
\ref{suthsh}.

Note that all variables $q^\alpha(t)$, $p^\alpha(t)$, and
$\rho_j^{\alpha\beta}(t)$ above were allowed to be complex. If we
restrict them to be real we recover the results in \cite{BL}. This
generalizes the results for the Sutherland model is Section
\ref{sutherland}. If we assume that all $q^\alpha(t)$ and
$p^\alpha(t)$ are purely imaginary and the $\rho_j^{\alpha\beta}(t)$
all real we obtain results generalizing the ones in Section
\ref{suthsh} (similarly as discussed in Section \ref{suthsh}, they can
be obtained from the ones for the real case by the substitutions in Eqs.\ 
(\ref{to1}) and (\ref{to2}).

\subsection{Example 5: Novel systems}
\label{general1}
The previous examples all where such that it was possible to give very
explicit solutions. As a final example we present a case where the
solution can be only described in a somewhat implicit way. It
corresponds to novel integrable spin-particle systems which
generalizes the model discussed in Section \ref{sec_simpex}. The
arguments here are similar to the ones in Section \ref{general}
and therefore we are rather brief.

We chose the charge distribution as follows
\begin{equation}
\label{rho2}
\rho^{\alpha\beta}(t,x) \equiv \frac{1}{\Delta_j
}e^{-ig[q^\alpha(t)-q^\beta(t)] x } s_j^{\alpha\beta}(t),\quad x_{j-1}\leq
x<x_j
\end{equation}
where $\Delta_j\equiv (x_j-x_{j-1})$, $j=1,\ldots,m$, and 
\begin{equation}  
 x_0\equiv -\pi < x_1 <
x_2< \ldots < x_{m-1}<x_{m} \equiv \pi 
\end{equation} 
with $m$ some positive
integer. This ansatz describes (essentially) piecewise constant
matter sources and is preserved under time evolution (cf.\
Eq. (\ref{3})). Moreover, using
\begin{equation}
\hat\rho^{\alpha\beta}(t,n)=\sum_{j=1}^m s_j^{\alpha\beta}(t)
\frac{1}{\Delta_j}\int_{x_{j-1}}^{x_j} dx\, e^{-ig[q^\alpha(t)-q^\beta(t)]x }
e^{-in x}
\end{equation}
we see that Eq.\ (\ref{rho_cons}) implies  the following constraint
\begin{equation}
\sum_{j=1}^m s_j^{\alpha\alpha}(0) = 0
\quad \forall \alpha=1,\ldots,N
\end{equation}
on possible initial conditions for the $s^{\alpha\beta}_{j}$.

Using the identities (\ref{id}) and (\ref{id1}), the Hamiltonian
(\ref{pq_hamiltonian}) becomes (again we set $a^\alpha(t)=0$ for
simplicity)
\begin{equation}
  \label{nov_ham1}
  {\cal H}=\frac{1}{2}\sum_{\alpha=1}^N (p^\alpha)^2  + \frac{1}{2}
  \sum_{ \stackrel{\alpha,\beta=1}{\alpha\neq \beta} }^N
  \sum_{j,k=1}^m
  w_{jk} (q^\alpha-q^\beta) s_k^{\alpha\beta} s_j^{\beta\alpha}
  +
  \frac{1}{2} \sum_{ \alpha=1}^N
  \sum_{j,k=1}^m
  d_{jk}s_k^{\alpha\alpha}s_j^{\alpha\alpha}
\end{equation}
where
\begin{eqnarray}
w_{jk} (r) &=& \frac{1}{\Delta_j\Delta_k}\int_{x_{j-1}}^{x_j}dx
\int_{x_{k-1}}^{x_k} dy
\,\frac{1}{4}e^{-i g r [(x-y)_{2\pi}-(x-y)] } \times 
\nonumber \\ \nopagebreak &\times&
\left(\frac{1}{\sin^2(\pi g r)} + \frac{i (x-y)_{2\pi}}{\pi} \cot(\pi g r)
-\frac{|(x-y)_{2\pi}|}{\pi}\right)
\end{eqnarray}
and
\begin{equation}
d_{jk} = \frac{1}{\Delta_j\Delta_k} \int_{x_{j-1}}^{x_j} dx
\int_{x_{k-1}}^{x_k}dy
\left(
\frac{(x-y)_{2\pi}^2}{8\pi^2} -
\frac{|(x-y)_{2\pi}|}{4\pi} + \frac{1}{12}\right)
\end{equation}
(of course these integrals could be further evaluated). The Poisson
brackets for this system are:
\begin{eqnarray}
  &\{p^\alpha,q^\beta\}& = \delta^{\alpha\beta}\\
  &\{s_j^{\alpha\beta},s_k^{\alpha'\beta'} \}& =
  ig 2\pi \delta_{jk}\left(
    \delta^{\beta\alpha'} s_j^{\alpha\beta'} -
    \delta^{\beta'\alpha} s_j^{\alpha'\beta} \right)
\end{eqnarray}
where the last equation is obtained by inserting
\begin{equation}  
\label{sr}
s_j^{\alpha\beta}(t)=\int_{x_{j-1}}^{x_j} dx \,
e^{ig[q^\alpha(t)-q^\beta(t)]x} \rho^{\alpha\beta}(t,x)
\end{equation}
in Eq.\ (\ref{rho_poi}).

We now show how to solve the initial value problem for the Hamiltonian
equations of this model. Writing Gauss' law in the form Eq.\ (\ref{s22})
and introducing ($Q_0=Q(0)$, as in the previous Section) 
$
B(0,x)= e^{ig Q_0 x} 
E(0,x)e^{-ig Q_0 x}
$
we obtain with Eq.\ (\ref{rho2}),
\begin{equation}  
\label{B1}
B^{\alpha\beta}(0,x)= C_j^{\alpha\beta}(0) +
\frac{(x-x_{j-1})}{(x_j-x_{j-1})}s^{\alpha\beta}_j(0) \quad \mbox{ for
$x_{j-1}\leq x<x_j$} 
\end{equation} 
where 
\begin{equation}  
 C_j^{\alpha\beta}(0) =
B^{\alpha\beta}(0,-\pi) + \sum_{k=1}^{j-1} s_k^{\alpha\beta}(0) .
\end{equation} 
Similarly as in Section \ref{general}, the conditions 
$2\pi p^\alpha(0)= \int_{-\pi}^\pi dx \,
E^{\alpha\alpha}(0,x)$ 
and $E(0,\pi)
= E(0,-\pi)$ 
determine $B(0,-\pi)$, and we obtain after some
computations, 
\begin{eqnarray}
\label{B3}
B^{\alpha\beta}(0,-\pi) &=& \delta^{\alpha\beta}\left(p^\alpha(0) -
\sum_{j=1}^m \frac{\Delta_j}{2\pi} \left(\half s_j^{\alpha\alpha}(0) +
\sum_{k=1}^{j-1} s_k^{\alpha\alpha}(0)\right) \right) + \nonumber \\ 
\nopagebreak &+& 
\left(1-\delta^{\alpha\beta} \right)\frac{
e^{-ig[q^\alpha(0)-q^\beta(0)]\pi}
}{2i\sin(\pi
g[q^\alpha(0)-q^\beta(0)])} \sum_{j=1}^{m} s^{\alpha\beta}_j(0) .
\end{eqnarray}
 We thus obtain $A_1(t,x) = e^{-ig Q_0 x} \left[Q_0 +B(0,x)\right]
e^{ig Q_0 x}$.

To find the gauge transformation bringing this to the diagonal Coulomb
gauge we define $\tilde S(t,x)= e^{ig Q_0 x} S(t,x)$
converting the differential equation in 
(\ref{S_def}) to $\partial_1\tilde S(t,x)+igB(0,x)t\tilde S(t,x) $.
Requiring $S(t,-\pi)=I$ we obtain 
\begin{equation}  
 S(t,x)= e^{-ig Q_0 x}
T_j\left(t,\mbox{$\frac{(x-x_{j-1})}{(x_j-x_{j-1})}$}\right)
T_{j-1}(t,1)\cdots T_1(t,1) e^{-ig Q_0 \pi} \quad \mbox{for $x_{j-1}\leq
x<x_j$} 
\end{equation}
where 
\begin{equation}  
 T_j(t,r) = \cP\exp\left(- igt\Delta_j
\int_0^r d\xi[C_j(0) +\xi s_j(0)] \right) . 
\end{equation} 
This gives the solution of the model: 
The
$q^\alpha(t)$ are determined by the eigenvalues of $S(t,\pi)$ as in
the previous examples. To compute $s_j(t)$ is more complicated here,
however: One can use Eq.\ (\ref{sr}) with
$\rho(t,x)=U^{-1}(t,x)\rho(0,x)U(t,x)$ and $U(t,x)$ determined by
$S(t,x)$ as discussed in Section \ref{gauges}. In the present case
we could not make the solution more explicit.

It is possible to write down explicitly the Lax pair family for the
present model, and there is also a 1-parameter
family of conservation laws here, namely $tr[B(t,x)^n]$, $n\in\N$ and 
$x\in [-\pi,\pi]$, where
$B(t,x)$ is as in Eqs.\ (\ref{B1})--(\ref{B3}) with $q^\alpha(0)$,
$p^\alpha(0)$ and $s_j(0)$ replaced by $q^\alpha(t)$, $p^\alpha(t)$
and $s_j(t)$.

\appendix
\section{Appendix: Proofs }
\label{derivation}
In this Appendix we prove the identities Eqs. (\ref{id}) and (\ref{id1}).
To prove Eq.\ (\ref{id}) we define the following function,
\begin{equation}
g(r,s):= \sum_{n\in\Z}\frac{e^{i(n+r)s} }{(n+r)^2}
\label{f_def}
\end{equation}
($r$ is a complex and $s$ a variable).
We observe
$$
  \frac{\partial^2 g(r,s)}{\partial s^2} = -\sum_{n\in {\bf Z}}e^{i
  (n+r)s} = -2\pi\delta(s) \quad \mbox{ for $-\pi\leq s\leq \pi$} .
$$
Integrating this twice gives
\begin{equation}  
g(r,s) = A(r)+B(r)s - \pi|s| \quad \mbox{ for $-\pi\leq s\leq \pi$}
 \label{f_func}
\end{equation}
where $A(r)$ and $B(r)$ are integration constants. We now determine
$A(r)$ and $B(r)$. Using Eq.\ (\ref{id0}) we conclude that $$ g(r,0) =
A(r) = \sum_{n\in {\bf Z}} \frac{1}{(n+r)^2} = \frac{\pi^2}{\sin^2{\pi
r}} .
$$ Moreover, the definition Eq.\ (\ref{f_def}) of
$g(r,s)$ implies
$$
g(r,-\pi)e^{ir\pi} = g(r,\pi)e^{-ir\pi} .
$$
Inserting Eq.\ (\ref{f_func}) we get
$$
  B(r) = i\pi\cot{(\pi r)}.
$$
We thus obtain,
\begin{equation}  
\sum_{n\in\Z}\frac{e^{i ns} }{(n+r)^2} =
e^{-irs}\left( \frac{\pi^2}{\sin^2{\pi r}} + i\pi s\cot{(\pi r)} -\pi|s|
\right) \quad
\mbox{ for $-\pi\leq s\leq \pi$}.
\end{equation}
The l.h.s. of this equation obviously is periodic in $s$ with period
$2\pi$. To extend the r.h.s. to all real $s$ we therefore only need
to replace $s$ by $s_{2\pi} : = s - 2\pi n$
with the integer $n$ chosen such that $-\pi\leq s - 2\pi n <\pi$.
This proves Eq.\ (\ref{id}).

To derive Eq.\ (\ref{id1}) we define
\begin{equation}
  h(s):=\sum_{\stackrel{n\in\Z}{n \neq 0}}
  \frac{e^{ins}}{n^2}
\end{equation}
($s$ is real) and compute
$$
  \frac{d^2 h(s)}{d s^2} = -\sum_{\stackrel{n\in {\bf Z}}{n\neq 0}}
  e^{ins} =
  -2\pi \delta(s) +1 \quad \mbox{ for $-\pi\leq s\leq \pi$.}
$$
Integrating this twice yields
$$
  h(s)=-\pi |s|+\frac{s^2}{2}+Cs+D 
$$
where $C$ and $D$ are integration constants. Periodicity of $h(s)$ implies
$C=0$, and the formula
$$
  \sum_{n=1}^\infty \frac{1}{n^2} = \frac{\pi^2}{6}
$$
(see, e.g.,\ \cite{Grad}) determines $D=\frac{\pi^2}{3}$. Thus
\begin{equation}  
\sum_{\stackrel{n\in\Z}{n \neq 0}}
  \frac{e^{ins}}{n^2} = -\pi |s|+\frac{s^2}{2}+\frac{\pi^2}{3}
  \quad \mbox{ for
  $-\pi\leq s\leq \pi$}.
\end{equation}
Again the r.h.s. can be extended to all real $s$ by replacing $s$
by $s_{2\pi}$ as defined above. This proves Eq.\ (\ref{id1}).

\section{Appendix: Poisson bracket for the charges}
\label{sec_pb}
In this appendix we derive the Poisson bracket for the charges
$\rho(x)$, i.e., Eq. (\ref{rho_poi}).
As mentioned in the text, consistency of Gauss' law $G(x) \simeq 0$
requires
\begin{equation}
  \partial_0 G^{\alpha' \beta'}(y) = \{ G^{\alpha' \beta'}(y), H \}
  \simeq 0 \quad  \forall
  \alpha',\beta'=1,\ldots,N.
  \label{Gdot}
\end{equation}
Inserting Eqs.\ (\ref{hamiltonian}) and (\ref{Gauss}) we can rewrite
$\{ G^{\alpha' \beta'}(y), H \} \simeq 0$ as
\begin{eqnarray*}
  &&\frac{1}{2\pi} \int_{-\pi}^{\pi} dx\,
  \sum_{\alpha,\beta=1}^N A_0^{\beta\alpha}(x)
  \left\{ \rho^{\alpha\beta}(x),
  \rho^{\alpha'\beta'}(y) \right\}  \simeq \nonumber\\
  &\simeq& \frac{1}{2\pi}\int_{-\pi}^{\pi} dx\, \sum_{\alpha,\beta,\gamma=1}^N
ig  \left( \half
  \left\{ E^{\beta\alpha}(x)E^{\alpha\beta}(x),
    A_1^{\alpha'\gamma}(y)E^{\gamma\beta'}(y) -
  E^{\alpha'\gamma}(y)A_1^{\gamma\beta'}(y) \right\} \right. \nonumber\\
  &-& A_0^{\beta\alpha}(x)
  \left\{\partial_xE^{\alpha\beta}(x),
    A_1^{\alpha'\gamma}(y)E^{\gamma\beta'}(y) -
  E^{\alpha'\gamma}(y)A_1^{\gamma\beta'}(y)\right\} \nonumber\\
  &-& A_0^{\beta\alpha}(x)
  \left\{A_1^{\alpha\gamma}(x)E^{\gamma\beta}(x) -
  E^{\alpha\gamma}(x)A_1^{\gamma\beta}(x), \partial_y
  E^{\alpha'\beta'}(y)\right\}
  - ig A_0^{\beta\alpha}(x)  \nonumber\\ &&\times\left.
  \left\{A_1^{\alpha\gamma}(x)E^{\gamma\beta}(x) -
  E^{\alpha\gamma}(x)A_1^{\gamma\beta}(x),
  A_1^{\alpha'\gamma}(y)E^{\gamma\beta'}(y) -
  E^{\alpha'\gamma}(y)A_1^{\gamma\beta'}(y) \right\} \right).
\end{eqnarray*}
Here we used that the only non-vanishing
Poisson brackets are those for $A_1$ with $E$ and for $\rho$ with
itself. Using now the Poisson bracket (\ref{canon_rel})
the r.h.s. of this becomes
\begin{eqnarray*}
 &&\frac{1}{2\pi} \int_{-\pi}^{\pi}dx\,  \sum_{\alpha,\beta=1}^N
 A_0^{\beta\alpha}(x)\{\rho^{\alpha\beta}(x),
 \rho^{\alpha'\beta'}(y) \} =  \nonumber\\
 &=&
  0 + ig\int_{-\pi}^{\pi} dx  \sum_{\gamma=1}^N \left(
  A_0^{\alpha'\gamma}(x)E^{\gamma\beta'}(y)\partial_x\delta(x-y) -
  A_0^{\gamma\beta'}(x)E^{\alpha'\gamma}(y)\partial_x\delta(x-y)
  \right.
  \nonumber\\
&+&  \left. A_0^{\gamma\beta'}(x)E^{\alpha'\gamma}(x)\partial_x\delta(x-y) 
- A_0^{\alpha'\gamma}(x)E^{\gamma\beta'}(x)\partial_x\delta(x-y)
  \right)
  \nonumber\\
  &-& (ig)^2 \int_{-\pi}^{\pi} dx\, \delta(x-y) \left(
  \left[\left[A_1(y),A_0(y)\right],E(y)\right]^{\alpha'\beta'} +
      \left[\left[A_0(y),E(y)\right],A_1(y)\right]^{\alpha'\beta'}
      \right)
      \nonumber\\
  &=& ig\int_{-\pi}^{\pi} dx\,  \delta(x-y) \left(-\left[\partial_1
      E(y),A_0(x)\right]^{\alpha'\beta'}
  -ig \left[\left[A_1(y),E(y)\right],A_0(x)\right]^{\alpha'\beta'}
\right)
    \nonumber\\
  &=& (-ig) \int_{-\pi}^{\pi}dx\,  \delta(x-y) \left[\partial_1
  E(y)+ ig\left[A_1(y),E(y)\right],A_0(x)\right]^{\alpha'\beta'}  \nonumber\\
  &\simeq& (-ig) \int_{-\pi}^{\pi}dx\,  \delta(x-y)
  \left[\rho(y),A_0(y)\right]^{\alpha'\beta'} \nonumber\\
  &=&
\frac{1}{2\pi}
  \int_{-\pi}^{\pi} dx  \sum_{\alpha,\beta = 1}^N
  A_0^{\beta\alpha}(y) \left(ig 2\pi \left[
  \rho^{\alpha \beta'}(y) \delta^{\alpha'\beta} -
   \rho^{\alpha'\beta}(y)\delta^{\alpha \beta'} \right]
 \delta(x-y)\right).
\end{eqnarray*}
Here the second equality follows by using partial integration
and the Jacobi identity, and in the fourth equality Gauss'
law was used. From this the Poisson
brackets in  Eq.\ (\ref{rho_poi}) follow.

\end{document}